\documentclass[onecolumn, journal]{IEEEtran}
\usepackage{graphicx}
\usepackage{booktabs}
\usepackage{array}
\usepackage{cite}
\usepackage{color}
\usepackage{enumerate}
\usepackage{tabularx}
\usepackage{amsmath}
\usepackage{multicol}
\usepackage{nomencl}
\usepackage{float}
\usepackage{subfigure}
\usepackage{stfloats}
\usepackage{amsfonts}
\usepackage{amsthm}
\usepackage{amssymb}
\usepackage{autobreak}
\usepackage{arydshln}
\usepackage{algorithm}%,algpseudocode}
\usepackage{algorithmic}
\newcommand{\subparagraph}{}
\usepackage[explicit]{titlesec}
\usepackage[tone,extra,safe]{tipa}
\usepackage[colorlinks,linkcolor=black,citecolor=black]{hyperref}

\titlespacing*{\section}{0pt}{1.2ex plus .0ex minus .0ex}{.3ex plus .0ex}
\titlespacing*{\subsection}{0pt}{1.2ex plus .0ex minus .0ex}{.3ex plus .0ex}
\ifCLASSINFOpdf
\else
\fi
\begin{document}
\title{Distributed Frequency Restoration and SoC Balancing Control for AC Microgrids}
\author{Chang Yu, Xiaoqing Lu, Jingang Lai and Li Chai,
\thanks{This work was supported by the National Natural Science
Foundation of China under Grant 61773158; by the Fundamental Research
Funds for the Central Universities under Grant 2042019kf0186; by the Natural Science
Foundation of Hunan Province under Grant 2018JJ2051; by the Humboldt
Research Council under Grant HB1807005.}
\thanks{C. Yu and L. Chai are with the School of Information Science and Engineering, Wuhan University of Science and Technology, Wuhan 430081, PR China (email: yuchang@wust.edu.cn; chaili@wust.edu.cn).}
\thanks{X. Lu is with the School of Electrical Engineering and Automation, Wuhan University, Wuhan 430072, PR China (email: yuchang@whu.edu.cn; hzhouwuhee@whu.edu.cn and luxq@whu.edu.cn).}
\thanks{J. Lai is with the E.ON Energy Research Center, RWTH Aachen University,
Aachen 52074, Germany (e-mail: jinganglai@126.com).}
}
\maketitle
\begin{abstract}
This paper develops an improved distributed finite-time control algorithm for multiagent-based ac microgrids with battery energy storage systems (BESSs) utilizing a low-width communication network. The proposed control algorithm can simultaneously coordinate BESSs to eliminate any deviation from the nominal frequency as well as solving state of charge (SoC) balancing problem. The stability of the proposed control algorithm is established using Lyapunov method and homogeneous approximation theory, which guarantees an accelerated convergence within a settling time that does not dependent on initial conditions. Based on this, to significantly reduce the communication burdens, an event-triggered communication mechanism is designed which can also avoid Zeno behavior. Then sufficient conditions on the event-triggered boundary are derived to guarantee the stability and reliability of the whole systems. Practical local constraints are imposed to implement the control protocol, and the theoretical results are applied to a test system consisting of five DGs and five BESSs, which verifies the effectiveness of the proposed strategy.
\end{abstract}
\IEEEpeerreviewmaketitle
\section{Introduction}
\IEEEPARstart {I}n islanded microgrids (MGs), the intermittent characteristics of most DERs may cause load perturbations and significantly affect the system frequency in relatively small ac microgrids. Thus, the control of a microgrid is essential to improve the frequency synchronization performance. However, the primary droop controllers, locally implemented at each distributed DG unit, may lead to a steady-state deviation from the nominal frequency value due to the power mismatch \cite{simpson2013synchronization}. To accommodate this, BESSs in an islanded ac MG can enhance the system stability and reliability through being charged by DERs or discharged for peak shaving or supporting local loads during grid failures and electrical shortages \cite{3}. The BESSs are therefore indispensable modules that can buffer short-term power imbalances among DERs and loads \cite{4}. At the same time, the state of charge (SoC) balancing problem which may result in the overcharging and overdischarging actions of BESSs arises \cite{17,18,19}. However, a few works focus on solving the frequency restoration and SoC balancing problems at once.
%\begin{figure}[!t]
%\centering
%\includegraphics[width=0.95\columnwidth,height=4.8cm]{microgrid.eps}
%\caption{Schematic illustration of an ac microgrid. The red dotted lines between BESSs represent communication links.}
%\end{figure}

{\color{blue}Coordination of BESSs in the microgrid can be regarded as a multi agent system (MAS), then the distributed control structure that only requires information exchange among neighboring components through a local communication network is used to implement the secondary control. At present, many works have made a lot of valuable achievements in the secondary control of microgrid from their own perspectives, such as multi-agent cooperation \cite{31,41,4.1,4.2} and cyber-security \cite{4.3,4.4}. As the number of DGs increases, the degree of each agent may increase accordingly. One may inevitably suffer from the band-width constraints. It would be desirable to reduce the communication burdens between agents in microgrids. To solve this problem, the event-triggered control strategy maybe a feasible scheme for the MG distributed control systems. Under
such an event-triggered mechanism, the information exchange
is only needed when any event is triggered \cite{5,6,7,7.1,7.2}.}

Safety-critical loads in a microgrid require operation at the nominal frequency, e.g., 50 or 60 Hz. Obviously, the synchronization rate of a microgrid is severely reduced as the number of units in the network increases. And besides, the low inertia of DERs make the system sensitive to load changes, which enforce the system to frequently stay in abnormal frequency intervals for long periods of time. The traditional droop control provides a basic method to regulate the synchronization rate. If the slope of the droop curve, droop coefficient, are decreased, a good synchronization rate is obtained at the expense of degrading the frequency regulation, which is acceptable if the frequency deviations are within 2\% \cite{8}. Despite more convenient to regulate, this inherent tradeoff of the frequency-droop based system restricts the selection of droop coefficients. Therefore, to avoid the serious limitation of droop coefficients in terms of synchronization rate optimization, it is of paramount value to accelerate the synchronization process and improve the convergence of the frequency control.

{\color{blue}Since the frequency of a MG could be regulated by BESSs, the convergence speed of the controller for BESSs is essential for the frequency synchronization rate of the system. In general, existing works to increase the convergence speed are mainly through two kinds of strategies: 1) enlarging the coupling strength, optimizing communication weights, and designing optimal network topology \cite{9,10,11}; and 2) finite-time control techniques is investigated to optimize convergence time in \cite{12,13,14}, and several conditions are proposed to ensure the finite-time stability \cite{14.1}. For the first strategy, optimizing control parameters are often at the expense of system dynamic performance; the global topology of the network is always unknown to each distributed agent, and it's difficult to maintain the maximum entropy due to the plug and play characteristic of a practical microgrid; besides, the network topology dynamically changes
randomly, and it cannot change according to the trend that
is beneficial to increase the algebraic connectivity. Therefore,
changing network topology is not the best choice to improve
the convergence rate and does not meet the actual applications
requirements\cite{14.2}. Moreover, the strategies mostly focus on methods that only achieve asymptotic consensus, which implies that convergence rate is at best exponential with an infinite settling time. Using a finite-time control, the consensus can be reached in finite time, and the closed-loop systems usually demonstrate better disturbance rejection properties. However, in the above mentioned studies related to finite-time consensus protocols, the estimation of the convergence time may be very conservative because of the large initial states. In addition, we cannot obtain an explicit estimation of the settling time before running the system. Moreover, complex operating conditions in an ac MG may lead to a big perturbation of initial conditions of the control system, which may cause a conservative convergence time accordingly \cite{15,16}. To this end, this paper mainly concerns with solving the frequency restoration and SoC balancing problems in a short settling time period that doesn't change with the initial conditions as well, by designing an improved finite-time event-triggered control. The main contributions lie in:}

1) Unlike most existing distributed cooperative schemes, which separately investigate frequency restoration and SoC balancing problems \cite{simpson2013synchronization,17}, the proposed control scheme is implemented through the secondary control layer in which the mentioned two problems can be solved simultaneously. Moreover, the secondary control proposed in this paper has a time scale that is independent of the primary control, which addresses the disadvantage that the primary and secondary control loops may interact in an adverse way \cite{dorfler2016breaking}.

2) Compared with the most existing finite-time controllers proposed in \cite{12,13,14}, an improved finite-time controller is designed to solve the above two problems within a short settling time regardless of the variation of the initial condition by combining the Lyapunov and homogeneous approximation theory.

3) Under the proposed event-triggered mechanism, each BESS only needs to update its own control input at event times using the local and neighboring discrete information, which can then effectively avoid global information collection and continuous signal sampling. This significantly reduces communication burdens. Further, the inter-event time is proved to be greater than zero so that the Zeno behavior can be effectively avoided.

The remain parts are organized as follows. We review notation, fundamental algebraic graph theory and homogeneity. Then, the aforementioned problems are formulated in Section II. In Section III, we present the proposed control algorithm, establish its stability, and design a control scheme for implementing the proposed algorithm. In Section IV, we provide the simulation results for an ac microgrid. Finally, Section V concludes this paper.
\section{Preliminaries}
Figure 1 depicts the cyber-physical layout of an inverter-based ac microgid augmented with BESSs. Based on a MAS
framework, an agent is assigned to each BESS in the microgrid, and exchanges its information with a few neighbors on a sparse and bidirectional communication graph.

\emph{A. Notation, Algebraic Graph Theory, and Homogeneity}

The following notation for vectors and matrices will be used throughout this paper. $\mathbb{R}_+$ denotes the set $[0,+\infty)$. Given a $n$-tuple $\left( x_1,\cdots,x_n \right)  $, let $ \underline{x} \in\mathbb{R}^n $ be the associated vector. Let $\boldsymbol{1}_n\in \mathbb{R}^n$ and $\boldsymbol{0}_n\in \mathbb{R}^n$ be the $n$-dimensional vectors of unit and zero entries. Let $\boldsymbol{J}_n\in \mathbb{R}^{n\times n}$ be the $n\times n$ matrix with all elements one, and $\boldsymbol{O}_n\in \mathbb{R}^{n\times n}$ be the $n\times n$ matrix with all elements zero. Let $\boldsymbol{1}_{n}^{\bot}=\left\{ \underline{x}\in \mathbb{R}^n:\boldsymbol{1}_n^T\underline{x}=0 \right\} $ be the orthogonal complement of $\boldsymbol{1}_n$. We denote the identity matrix by $\boldsymbol{I}_n\in \mathbb{R}^{n\times n}$. Given an ordered index set $\mathcal{N}$ and a $1$-dimensional array $\left\{ x_i \right\} _{i\in \mathcal{N}}$, let $\operatorname{diag}\left( \left\{ x_i \right\} _{i\in \mathcal{N}} \right) \in \mathbb{R}^{\left| \mathcal{N} \right|\times \left| \mathcal{N} \right|}$ be the associated diagonal matrix. Let $\lVert \cdot  \rVert $ denote the Euclidean norm. For a matrix $A\in ^{n\times n}$, let $\lambda_1(A) \leq \cdots \leq \lambda_n(A)$ be its eigenvalues. Given a vector $\underline{x} \in \mathbb{R}^n$, define $\operatorname{sig}(\underline{x})^\gamma=[|x_1|^\gamma\text{sgn}(x_1),\cdots,|x_n|^\gamma\text{sgn}(x_n)]^T$ and
$|\underline{x}|^{\gamma}=\left[\left|x_{1}\right|^{\gamma}, \ldots,\left|x_{n}\right|^{\gamma}\right]^{T}$ where $\text{sgn}(\cdot)$ is the signum function. The function $\mathfrak{H}:\mathbb{R}_+^2\to \mathbb{R}_+$ is defined as $\mathfrak{H}(a,b)=\frac{a}{1+a}(1+b)$. Given $\underline{r}\in \mathbb{R}^n$ and $\varepsilon \in \mathbb{R}$, $\varepsilon^{\underline{r}} \diamond \underline{x}=\left(\lambda^{r_{1}} x_{1}, \ldots, \lambda^{r_{n}} x_{n}\right)^{T}$ is
the dilation of a vector $\underline{x}$ in $\mathbb{R}^n$ with weight $\underline{r}$.

We denote by $\mathcal{G} \left( \mathcal{V} ,\mathcal{E} ,A \right) $ an undirected, connected and simple graph, where $\nu =\left\{ \text{1,}\cdots ,n \right\} $ is the set of vertexes, $\mathcal{E} \subset \mathcal{V} \times \mathcal{V} $ is the set of edges, and $A\in \mathbb{R}^{n\times n}$ is the adjacency matrix. The entries of $A$ satisfy $a_{ij}=1$ for each undirected edge $e=\left( i,j \right) \in \mathcal{E} $ and $a_{ij}=0$ otherwise. The set of neighbors of node $i$ is denoted as $\mathcal{N}_i=\left\{j |\left({i}, {j}\right) \in \mathcal{E}\right\}$. For each node $i\in \mathcal{V}$, we define the degree by $\text{d}_i=\sum_{j=1}^n{a_{ij}}$, and the associated degree matrix $D=\operatorname{diag}\left( \left\{ \text{d}_i \right\} \right) \in \mathbb{R}^{n\times n}$. The Laplacian matrix is defined as $L=D-A$.

A continuous vector function $f:\mathbb{R}^n \to \mathbb{R}^n$ is said homogeneous in the $l$-limit with associated triple $\left(\underline{r}_{l}, k_{l}, \phi_{l}\right)$, where $\underline{r}_{l}$ in $i n\left(\mathbb{R}_{+} \backslash\{0\}\right)^{n}$ is the weight, $d_l$ in $\mathbb{R}$ is the degree and $f_l$ is the approximating vector function, if, for each $i$ in $\{1, \ldots, n\}$, if, for each $i$, $d_{l}+r_{l, i}>0$ and the condition $\lim _{\varepsilon \rightarrow l} \max _{x \in C}|\frac{f_i\left(\varepsilon^{\underline{r}_{l}} \diamond \underline{x}\right)}{\varepsilon^{d_{l}+r_{l, i}}}-f_{l,i}(\underline{x})|=0$ holds for each compact set $C$ in $\mathbb{R}^{n} \backslash\{0\}$ \cite{171}.
\begin{figure}[!t]
\centering
\includegraphics[width=0.5\columnwidth]{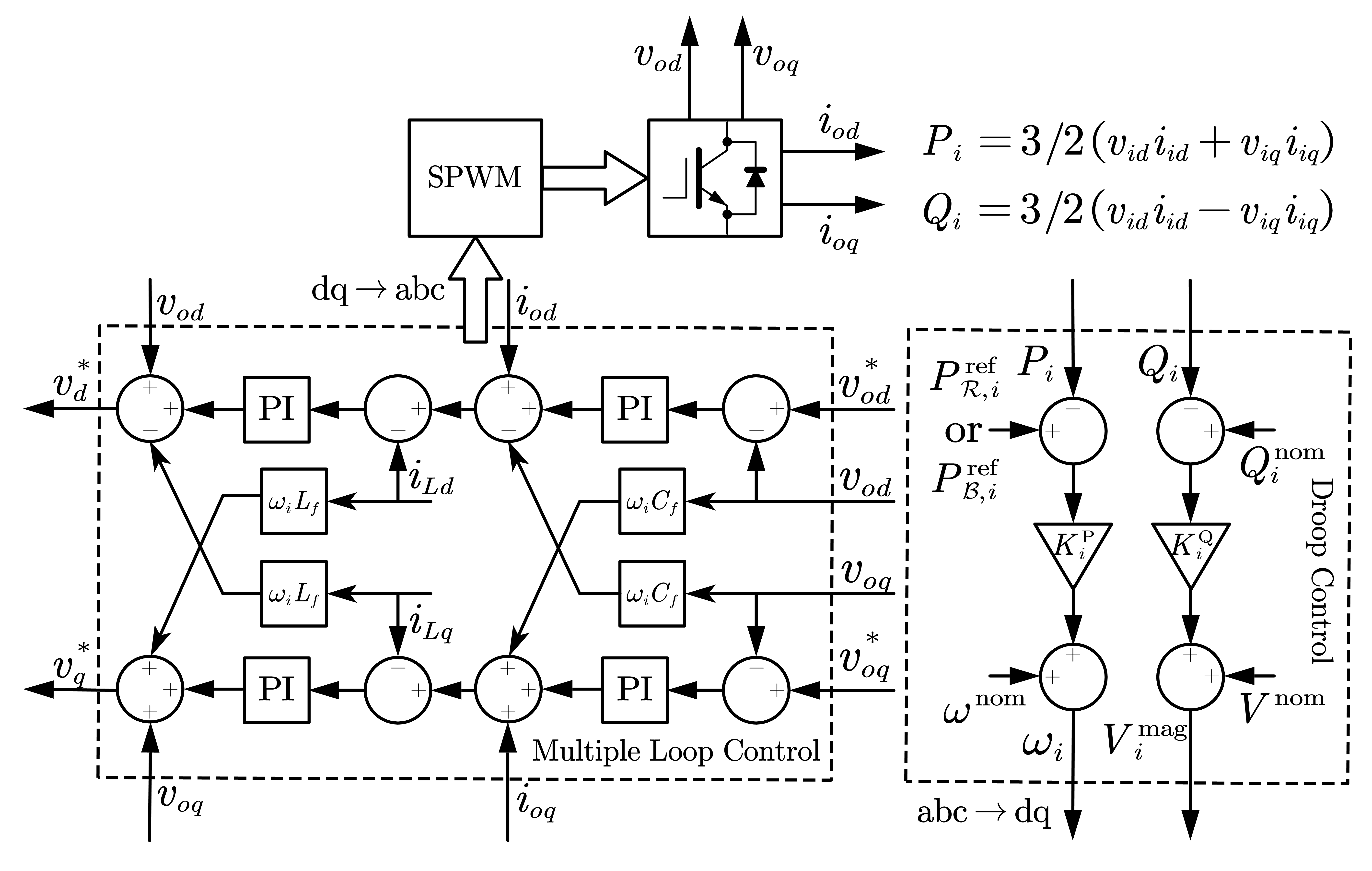}
\caption{The internal multiple loop control diagram for inverters injecting power into the network.}\label{fig:1}
\end{figure}

\emph{B. Dynamic Modeling of Inverter-based AC Microgrids}

{\color{blue}Consider a lossless inverter-based ac microgrid operating in islanded mode. The physical system of the microgrid is modeled as an undirected, connected and simple graph $\mathcal{G}_p\left( \mathcal{V} ,\mathcal{E} \right) $, where $\mathcal{V}$ is the set of nodes and $\mathcal{E}$ is the set of power lines connecting those nodes. A subset $\mathcal{R}\in \mathcal{V}$ represents the inverters interfaced with uncontrollable renewable RESs such as PVs that is operating under the MPPT mode. Another subset $\mathcal{B}$ represents the bidirectional inverters through which BESSs could operate in charging/discharging mode. In addition, we denote by $\mathcal{L} :=\mathcal{V} \backslash (\mathcal{R}\cup \mathcal{B})$ the set of loads. To each inverter $i\in \mathcal{R}\cup \mathcal{B}$, we assign a voltage signal of the sinusoidal function form $V_i(t) = \sqrt{2}V_i^\mathrm{mag} \cos(\omega^\mathrm{nom}t+\varphi_i)$ or the phase vector $V_i^\mathrm{mag}\angle\varphi_i$, where $V_i^\mathrm{mag}$ is the nodal voltage amplitude, $\omega^\mathrm{nom}$ is the nominal frequency that is equal to $2 \pi \cdot 50$ or $2 \pi \cdot 60$ Hz, and $\varphi_i$ is the nodal voltage phase angle. The angular frequency of inverter $i$ is denoted by $\omega_i$. The active and reactive electrical power injected into or absorbed from the network at inverter $i$ is denoted by $P_i$ and $Q_i$. For $i\in \mathcal{R}$, the uncontrollable active power injection $P_i$ is restricted to the interval $[0,P_{\mathcal{R},i}^\mathrm{rat}]$ where $P_{\mathcal{R},i}^\mathrm{rat}>0$ is the rating of inverter $i$. For $i\in \mathcal{B}$, the charging/discharging power $P_i$ is positive when inverter $i$ is in charging mode and negative when in discharging mode, and it is restricted to the interval $[P_{\mathcal{B},i}^\mathrm{cha},P_{\mathcal{B},i}^\mathrm{dis}]$ where $P_{\mathcal{B},i}^\mathrm{dis}>0$ and $-P_{\mathcal{B},i}^\mathrm{cha}>0$ are respectively the upper bounds of discharging power and charging power of inverter $i$. For $(i,j)\in \mathcal{E}$, $Z_{ij}\angle\delta_{ij}$ is the impedance of the power line between node $i$ and $j$. Since the outpower of RESs is intermittent, the BESSs are even more important for the power balance of the system such that it can shift electricity from peak periods to off peak periods. Therefore, we only consider the control of BESSs in this paper.}
\begin{figure}[!t]
\centering
\includegraphics[width=0.5\columnwidth]{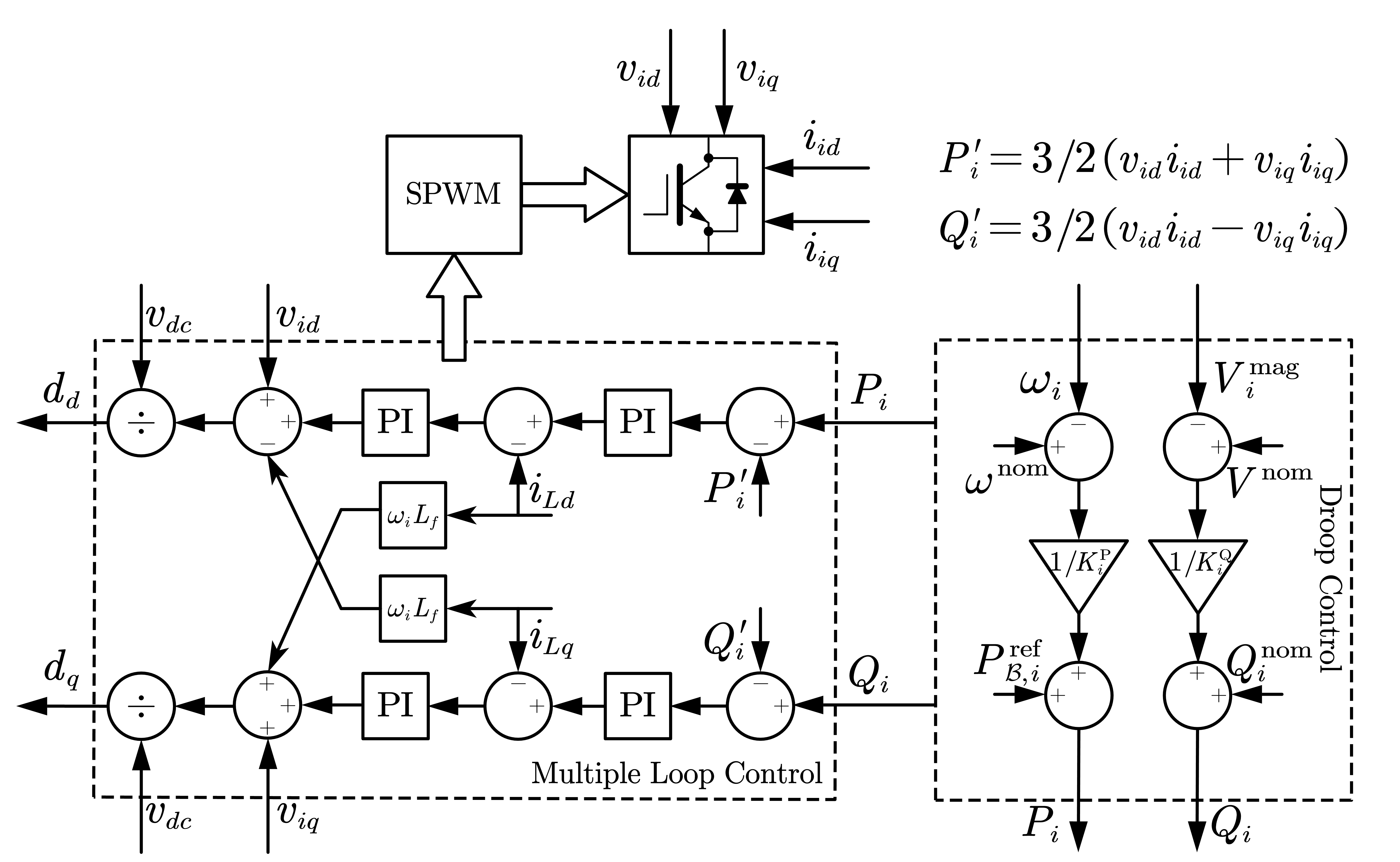}
\caption{The internal multiple loop control diagram for inverters absorbing power from the network.}\label{fig:2}
\end{figure}
Assume that the voltage of inverter $j$ that is adjacent to inverter $i$ is $V_j^\mathrm{mag}\angle\varphi_j$. Then, the complex power delivered from the $i$th to the $j$th inverter is
\begin{equation}\label{eq:1}
S_{ij}=V_{ij} I_{ij}^{*}=\frac{V_i^\mathrm{mag} V_j^\mathrm{mag} \angle\left(\delta_{ij}-\left(\varphi_{i}-\varphi_{j}\right)\right)}{Z_{ij}}-\frac{{V_i^\mathrm{mag}}^{2} \angle \delta_{ij}}{Z_{ij}}.
\end{equation}
From ({\ref{eq:1}}), the real and reactive powers from $i$th to $j$th inverter are obtained, respectively, as
\begin{equation}\label{eq:2}
\left\{\begin{array}{l}
{P_{ij}=\frac{V_i^\mathrm{mag} V_j^\mathrm{mag} \cos \left(\delta_{ij}-\left(\varphi_{i}-\varphi_{j}\right)\right)}{Z_{ij}}-\frac{{V_i^\mathrm{mag}}^{2} \cos \delta_{i}}{Z_{ij}}} \\
{Q_{ij}=\frac{V_i^\mathrm{mag} V_j^\mathrm{mag} \sin \left(\delta_{ij}-\left(\varphi_{i}-\varphi_{j}\right)\right)}{Z_{ij}}-\frac{{V_i^\mathrm{mag}}^{2} \sin \delta_{ij}}{Z_{ij}}}
\end{array}\right..
\end{equation}
If the resistance of the line is small enough to be omitted, i.e. $\delta_{ij}\to 90^{\deg}$, (\ref{eq:2}) can be reduced as
\begin{equation}\label{eq:3}
\left\{\begin{array}{l}
{P_{ij}=\frac{V_i^\mathrm{mag} V_j^\mathrm{mag} \sin \left(\left(\varphi_{i}-\varphi_{j}\right)\right)}{Z_{ij}}} \\
{Q_{ij}=\frac{V_i^\mathrm{mag} V_j^\mathrm{mag} \cos \left(\left(\varphi_{i}-\varphi_{j}\right)\right)}{Z_{ij}}-\frac{{V_i^\mathrm{mag}}^{2} }{Z_{ij}}}
\end{array}\right..
\end{equation}
Thus, the frequency and voltage droop characteristics can be utilized to regulate the inverter’s frequency and voltage via primary conventional droop mechanisms:
\\For $i\in \mathcal{R}$,
\begin{equation}\label{eq:4}
\left\{\begin{array}{l}
{\omega_i=\omega^{\mathrm{nom}}-K_i^\mathrm{P} (P_i-
P_{\mathcal{R},i}^{\mathrm{nom}})} \\
{V_i^\mathrm{mag}=V^{\mathrm{nom}}-K_i^\mathrm{Q} Q_i^{\mathrm{nom}}}
\end{array}\right.,
\end{equation}
\\for $i\in \mathcal{B}$,
\begin{equation}\label{eq:5}
\left\{\begin{array}{l}
{\omega_i=\omega^{\mathrm{nom}}-K_i^\mathrm{P} (P_i-
P_{\mathcal{B},i}^{\mathrm{ref}})} \\
{V_i^\mathrm{mag}=V^{\mathrm{nom}}-K_i^\mathrm{Q} Q_i^{\mathrm{nom}}}
\end{array}\right.,
\end{equation}
where the constant $P_{\mathcal{R},i}^{\mathrm{nom}}\in [0,P_{\mathcal{R},i}^\mathrm{rat}]$ is a nominal value that is relevant the power rating of the RES interfaced with inverter $i$, the control variable $P_{\mathcal{B},i}^{\mathrm{ref}}\in [P_{\mathcal{B},i}^\mathrm{cha},P_{\mathcal{B},i}^\mathrm{dis}]$ is the reference power of inverter $i$ that serves as an input to regulate the charging/discharging power $P_i$ and can be selected at the secondary control level. $K_i^Q$ and $K_i^P$ are the corresponding droop coefficients. It is worth noting that, for $i\in \mathcal{B}$, the $p-f$ droop equation in (\ref{eq:5}) is valid when inverter $i$ is in charging mode if $K_i^\mathrm{P}$ is negative. To implement the droop mechanisms for inverter $i$, local multiple loop decoupling schemes are adopted. Fig. \ref{fig:1} and Fig. \ref{fig:2} represent the multiple loop control diagrams when inverter $i$ is injecting active power into and absorbing active power from the network respectively. The local multiple loop decoupling scheme shown in Fig. \ref{fig:1} can make the output voltage of the inverter track its reference value in short time, whereas the one shown in Fig. \ref{fig:2} can make the input
active power track its reference value in short time. Obviously, the control scheme in Fig. \ref{fig:1} applies to inverters belonging to set $\mathcal{R}$ and discharging inverters belonging to set $\mathcal{B}$. The control scheme in Fig. \ref{fig:2} applies to the charging inverters belonging to set $\mathcal{B}$. Thus, for inverters belonging to set $\mathcal{B}$, the above two control schemes need to be switched momentarily according to the operation requirements of the corresponding BESSs. The details about the multiple loop control algorithms are ignored. If needed, one can refer to \cite{20}.

Since we mainly focus on the frequency regulation in this paper, the former part of (\ref{eq:4}) and (\ref{eq:5}), i.e. the $p-f$ droop control mechanism, will be utilized to establish the dynamics of the inverter-based ac microgrid. For simplicity, we denote by $\omega_i^\mathrm{dev}\triangleq \omega_i-\omega^\mathrm{nom}$ the nodal angular frequency deviation from the nominal angular frequency. Hence, by combining (\ref{eq:3})-(\ref{eq:5}), the network dynamics are modeled by the swing equations
\begin{subequations}\label{eq:6}
\begin{equation}
\frac{\omega_i^\mathrm{dev}}{K_i^\mathrm{P}}\!=\!P_{\mathcal{R},i}^{\mathrm{nom}}\!-\!\sum_{j\in \mathcal{N}_i}\frac{V_i^\mathrm{mag} V_j^\mathrm{mag} \sin \left(\varphi_{i}-\varphi_{j}\right)}{Z_{ij}}, i\in \mathcal{R}
\end{equation}
\begin{equation}
\frac{\omega_i^\mathrm{dev}}{K_i^\mathrm{P}}=P_{\mathcal{B},i}^{\mathrm{ref}}-\sum_{j\in \mathcal{N}_i}\frac{V_i^\mathrm{mag} V_j^\mathrm{mag} \sin \left(\varphi_{i}-\varphi_{j}\right)}{Z_{ij}}, i\in \mathcal{B}
\end{equation}
\begin{equation}
0=P_{\mathcal{L},i}-\sum_{j\in \mathcal{N}_i}\frac{V_i^\mathrm{mag} V_j^\mathrm{mag} \sin \left(\varphi_{i}-\varphi_{j}\right)}{Z_{ij}}, i\in \mathcal{L}
\end{equation}
\end{subequations}
where $P_{\mathcal{L},i}<0$ is the load demand at node $i\in \mathcal{L}$ and $P_L\triangleq -\sum_{i \in \mathcal{L}}P_{\mathcal{L},i}$ is the local estimation of total load demand at the virtual leader node $r\in \mathcal{B}$. To maintain the power balance of the system in the situation of time-varying load demand, the following constraints should be satisfied:
\begin{equation}\label{6.5}
\left\{\begin{array}{l}
{\max\left(P_L\right)- \sum_{i \in \mathcal{R}} P_{\mathcal{R},i}^{\mathrm{nom}}  \leq \sum_{i \in \mathcal{B}} P_{\mathcal{B},i}^{\mathrm{dis}}} \\
{{\min\left(P_L\right)- \sum_{i \in \mathcal{R}} P_{\mathcal{R},i}^{\mathrm{nom}}  \geq \sum_{i \in \mathcal{B}} P_{\mathcal{B},i}^{\mathrm{cha}}} }
\end{array}\right.,
\end{equation}
which means that the total charging/discharging power of BESSs is able to shift electricity from peak periods to off peak periods.

\emph{C. Finite-Time Frequency Restoration and SoC Balancing}

We are interested in frequency-synchronized solutions of model (\ref{eq:6}), satisfying $\omega_i^\mathrm{dev}=\omega^\mathrm{syn}$ for some $\omega^\mathrm{syn}\in \mathbb{R}$. Summing over equations (\ref{eq:6}) and evaluating $\omega_i^\mathrm{dev}=\omega^\mathrm{syn}$ for $i\in \mathcal{R}\cup \mathcal{B}$, we obtain the synchronization frequency
\begin{equation}\label{eq:7}
\omega_\mathrm{syn}=\frac{\sum_{i \in \mathcal{R}} P_{\mathcal{R},i}^{\mathrm{nom}}+ \sum_{i \in \mathcal{B}} P_{\mathcal{B},i}^{\mathrm{ref}}+\sum_{i \in \mathcal{L}}P_{\mathcal{L},i} }{\sum_{i\in \mathcal{R}\cup \mathcal{B}} \frac{1}{K_i^\mathrm{P}}},
\end{equation}
which implies that all the inverters operate under the nominal angular frequency in the steady state only if
\begin{equation}\label{eq:8}
\sum_{i \in \mathcal{B}} P_{\mathcal{B},i}^{\mathrm{ref}}=-\sum_{i \in \mathcal{R}} P_{\mathcal{R},i}^{\mathrm{nom}}+P_L,
\end{equation}
where the right side of equation (\ref{eq:8}) is the total power mismatch and denoted by $P^\mathrm{mis}$. If $P^\mathrm{mis}$ is equal to $0$, the synchronization angular frequency is equal to the nominal value $\omega^\mathrm{nom}$ accordingly. Thus, one can regulate the references $P_{\mathcal{B},i}^{\mathrm{ref}}$ to solve the frequency restoration problem.
\begin{figure}[t]
\centering
\includegraphics[width=0.5\columnwidth]{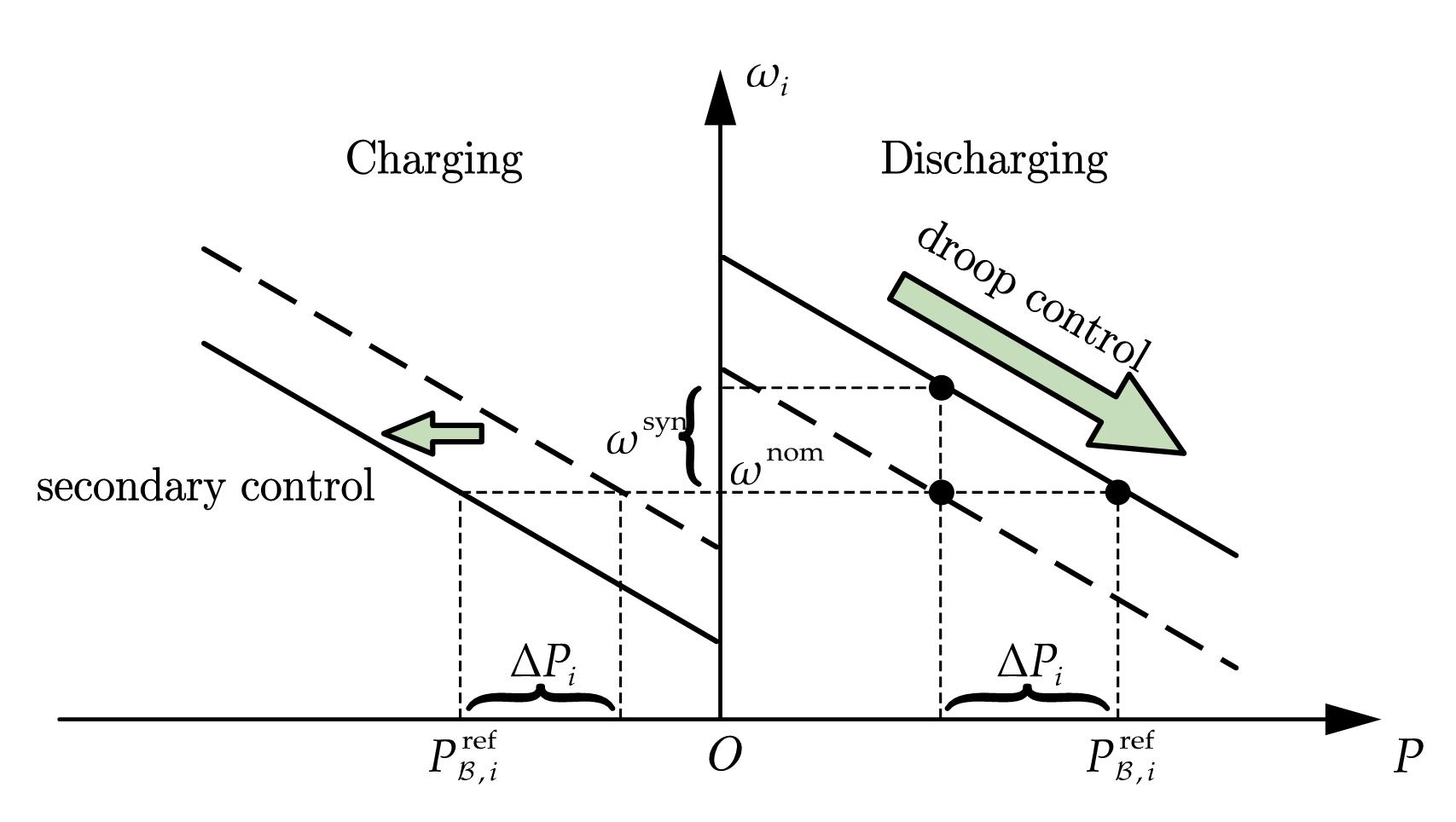}
\caption{Procedure of frequency regulation.}\label{fig:3}
\end{figure}
The procedure of frequency regulation is shown in Fig. \ref{fig:3}. Suppose all the inverters in the ac microgrid operate under the nominal angular frequency $\omega^\mathrm{nom}$ at $t_0$. Once total load $P_L$ change $\varDelta P_L$, then the steady-state angular frequency deviation $\omega^\mathrm{syn}$ will be forced to be nonzero due to the $p-f$ droop mechanism. Hereafter, the activated secondary control law will enforce each inverter $i\in \mathcal{B}$ to operate in charging/discharging mode according to the sign of $\varDelta P_L$, and regulate reference $P_{\mathcal{B},i}^{\mathrm{ref}}$ to compensate the variation $\varDelta P_L$. Hence, the secondary control finally maintain the restore the synchronization angular frequency to the nominal value.
\par While the above analysis gives one of the control objectives of frequency restoration, it offers no strategy on how to regulate the references $P_{\mathcal{B},i}^{\mathrm{ref}}$ to satisfy the actuation constraint (\ref{6.5}). Motivated by the power sharing strategy in \cite{simpson2013synchronization}, the following definition gives the proper criteria for regulation.

\textbf{\emph{Definition 1:}}
\rm The desired references $P_{\mathcal{B},i}^{\mathrm{ref}}$ share the total power mismatch proportionally if $P_{\mathcal{B},i}^{\mathrm{ref}}/P_{\mathcal{B},i}^\mathrm{cha}=P_{\mathcal{B},j}^{\mathrm{ref}}/P_{\mathcal{B},j}^\mathrm{cha}$ in charging mode and $P_{\mathcal{B},i}^{\mathrm{ref}}/P_{\mathcal{B},i}^\mathrm{dis}=P_{\mathcal{B},j}^{\mathrm{ref}}/P_{\mathcal{B},j}^\mathrm{dis}$ in discharging mode, for each $i$, $j\in \mathcal{B}$.

Similar to Theorem 7 in \cite{simpson2013synchronization}, one can prove that the actuation constraint (\ref{6.5}) is able to be satisfied if inverters belonging to $\mathcal{B}$ are regulated proportionally. Without loss of generality, the rest of this section discusses only the case when the inverters belonging to $\mathcal{B}$ are required to operate in discharging mode. The same principle can be applied when the inverters operate
in the charging mode. Note that $P_{\mathcal{B},i}^{\mathrm{ref}}/P_{\mathcal{B},i}^\mathrm{dis}$ autonomously
achieve an agreement for all $i\in \mathcal{B}$, we let $\lambda_i \triangleq P_{\mathcal{B},i}^{\mathrm{ref}}/P_{\mathcal{B},i}^\mathrm{dis}$ and $\lambda^*$ be the dynamic consensus value. Thus, it is easy to verify that
\begin{equation}\label{eq:8.5}
\lambda^*(t)=\frac{P^\mathrm{mis}(t)}{\sum_{i\in \mathcal{B}}P_{\mathcal{B},i}^\mathrm{dis}},
\end{equation}
which is known by the virtual leader.

Next, the SoC balancing problem is considered. An estimation of the SoC of the BESS interfaced with inverter $i\in \mathcal{B}$ can
be represented by the coulomb counting method as follows
\begin{equation}\label{eq:9}
\mathrm{SoC}_i=\mathrm{SoC}_{0}-\frac{1}{C_i} \int \eta_{i_i^\mathrm{dc}} \mathrm{d} t,
\end{equation}
where $\mathrm{SoC}_{0}$ is the initial value of the SoC, $C_i$ denotes the capacity of the BESS, $i_i^\mathrm{dc}$ is the output dc current of the BESS, and $\eta$ is the BESS discharging efficiency. Define $P_\mathrm{dc}$ and $v_i^\mathrm{dc}$ as the dc output power and voltage of the BESS. Assume the power loss of the interfacing inverter is negligible and $\omega_i=\omega^\mathrm{nom}$. Thus one has $P_{\mathcal{B},i}^{\mathrm{ref}}=P_i \approx  P_\mathrm{dc} = v_i^\mathrm{dc}i_i^\mathrm{dc}$ and $\eta\approx1$. Considering these assumptions, we derive the estimation of SoC with respect to the active power reference output power and initial SoC value, as
\begin{equation}\label{eq:10}
\mathrm{SoC}_i=\mathrm{SoC}_{0}-\frac{1}{C_i} \int \frac{P_{\mathcal{B},i}^{\mathrm{ref}}}{v_i^\mathrm{dc}} \mathrm{d} t.
\end{equation}
Then, by differentiating (\ref{eq:10}) with respect to time $t$, one can obtain
\begin{equation}\label{eq:11}
\dot{\mathrm{SoC}}_i=-\frac{1}{C_i} \frac{P_{\mathcal{B},i}^{\mathrm{ref}}}{v_i^\mathrm{dc}}=-\frac{1}{C_i} \frac{\lambda_iP_{\mathcal{B},i}^\mathrm{dis}}{v_i^\mathrm{dc}}.
\end{equation}
Assume that $K^\mathrm{SoC} \triangleq -{P_{\mathcal{B},i}^\mathrm{dis}}/{C_iv_i^\mathrm{dc}}$ is constant for all $i\in \mathcal{B}$, and hence, the dynamic of the virtual leader $r$ is modeled as following:
\begin{equation}
\left\{\begin{array}{l}
{\dot{\mathrm{SoC}}_r(t)=K^\mathrm{SoC}\lambda_r(t)} \\
{\dot{\lambda}_r(t)={\dot{P}^\mathrm{mis}(t)}/\left({\sum_{i\in \mathcal{B}}P_{\mathcal{B},i}^\mathrm{dis}}\right)}
\end{array}\right..
\end{equation}
Then, the frequency restoration and SoC balancing problem can be regarded as a cooperative tracker problem. In order to adjust state variables $\lambda_i$ and $\mathrm{SoC}_i$ of follower $i\in \mathcal{B}$ dynamically to track the trajectories of the virtual leader $r$, a secondary integral controller is inserted in front of it
\begin{equation}\label{eq:12}
\dot{\lambda_i}=u_i,
\end{equation}
where $u_i$ is the input signal of the inserted integrator controller and only requires information from neighboring agents of inverter $i$. Moreover, the secondary integral controllers $u_i$ are sought to solve the frequency restoration and SoC balancing problems within a finite time, if there exists a settling time, $T_v$, such that
\begin{equation}\label{eq:12.5}
\left\{\begin{array}{l}
{\lim _{t \rightarrow t_0+T_{f}} \lambda_{i}(t)=\lambda^*(t)} \\
{\lambda_{i}(t)=\lambda^*\,\forall t \geqslant t_0+T_{f}\,\forall i \in \mathcal{B}}\\
{\lim _{t \rightarrow t_0+T_{f}} \mathrm{SoC}_i(t)=\mathrm{SoC}_j(t)} \\
{\mathrm{SoC}_i(t)=\mathrm{SoC}_j(t) \,\forall t \geqslant t_0+T_{f}\,\forall i,j \in \mathcal{B}}
\end{array}\right.
\end{equation}
This ensures that all the $\lambda_i(t)$ and $\mathrm{SoC}_i(t)$ can precisely reach the consensus after $t=t_{0}+T_{f}$.
\section{Main Results}
All the BESSs in the ac microgrid can be regarded as agents in the MAS. Suppose those agents communicate with each other through a undirected and connected communication network associated with a simple graph $\mathcal{G}_c(\mathcal{B},\mathcal{E}, L) $. $L$ is the corresponding Laplacian matrix of graph $\mathcal{G}_c$. The pinning gain $g_r > 0 $ for leader agent $r\in \mathcal{B}$ with access to the reference value. $G=\operatorname{diag}\left( \left\{ g_i \right\} \right) \in \mathbb{R}^{|\mathcal{B}|\times |\mathcal{B}|}$ is the diagonal matrix of pinning gains. Letting $s_i=\mathrm{SoC}_i$ and $q_i=K^\mathrm{SoC}\lambda_i$, the dynamics of agent $i$ and the virtual leader $r$ can be rewritten as
\begin{equation}\label{eq:13}
\left\{\begin{array}{l}
{\dot{s}_{i}(t)=q_{i}(t)} \\
{\dot{q}_{i}(t)=u_{i}(t)}
\end{array}\right., i\in \mathcal{B},
\;
\left\{\begin{array}{l}
{\dot{s}_{r}(t)=q_{r}(t)} \\
{\dot{q}_{r}(t)=\mathcal{M}(t)}
\end{array}\right.,
\end{equation}
where $\mathcal{M}(t)=K^\mathrm{SoC}{\dot{P}^\mathrm{mis}(t)}/\left({\sum_{i\in \mathcal{B}}P_{\mathcal{B},i}^\mathrm{dis}}\right)$. Since the time-scale of the convergence for the secondary controller is much shorter than that of the total power mismatch, it is reasonable to assume that $q_r$ is a piecewise constant signal with successive variation time.

\emph{A. Distributed Finite-Time and Event-Driven Control}

We construct an improved distributed finite-time and event-triggered control law for the systems in (\ref{eq:13}) as follows
\begin{equation}\label{eq:14}
\begin{aligned}
&u_{i}(t)=u_{i1}(t)+u_{i2}(t),\\
&u_{i1}(t)=\\
&-k_1 \operatorname{sig}\bigg( \alpha\sum_{j\in\mathcal{N}_{i}}s_{i}\left(t_{k}^{i}\right)-s_{j}\left(t_{k}^{i}\right)+g_i\left(s_{i}\left(t_{k}^{i}\right)-s_{r}\left(t_{k}^{i}\right)\right)
\\&+\beta \sum_{j\in\mathcal{N}_{i}}q_{i}\left(t_{k}^{i}\right)-q_{j}\left(t_{k}^{i}\right)+g_i\left(q_{i}\left(t_{k}^{i}\right)-q_{r}\left(t_{k}^{i}\right)\right)\bigg)^{\gamma_1}\\
&-k_2 \operatorname{sig}\bigg(\alpha\sum_{j\in\mathcal{N}_{i}}s_{i}\left(t_{k}^{i}\right)-s_{j}\left(t_{k}^{i}\right)+g_i\left(s_{i}\left(t_{k}^{i}\right)-s_{r}\left(t_{k}^{i}\right)\right)
\\
&+\beta \sum_{j\in\mathcal{N}_{i}}q_{i}\left(t_{k}^{i}\right)-q_{j}\left(t_{k}^{i}\right)+g_i\left(q_{i}\left(t_{k}^{i}\right)-q_{r}\left(t_{k}^{i}\right)\right)\bigg)^{\gamma_2},\\
&u_{i2}(t)=\\
& -k_3\bigg\{ \alpha\sum_{j\in\mathcal{N}_{i}}s_{i}\left(t_{k}^{i}\right)-s_{j}\left(t_{k}^{i}\right)+g_i\left(s_{i}\left(t_{k}^{i}\right)-s_{r}\left(t_{k}^{i}\right)\right)
\\
&+\beta \sum_{j\in\mathcal{N}_{i}}q_{i}\left(t_{k}^{i}\right)-q_{j}\left(t_{k}^{i}\right)+g_i\left(q_{i}\left(t_{k}^{i}\right)-q_{r}\left(t_{k}^{i}\right)\right)\bigg\},\\
&t \in\left[t_{k}^{i}, t_{k+1}^{i}\right),
\end{aligned}
\end{equation}
where $t_{k}^{i}$ is the latest event instant of agent $i$. $0<\gamma_{1}<1$, $\gamma_{2}>1$, $k_1>0$, $k_2>0$, $k_3>0$, $\alpha>0$ and $\beta>0$ are positive constant coefficients. For brevity, we denote
\begin{equation*}
\xi_i\left(t_{k}^{i}\right)=\sum_{j\in\mathcal{N}_{i}}s_{i}\left(t_{k}^{i}\right)-s_{j}\left(t_{k}^{i}\right)\!+\!g_i\left(s_{r}\left(t_{k}^{i}\right)-s_{i}\left(t_{k}^{i}\right)\right),
\end{equation*}
\begin{equation*}
\zeta_i\left(t_{k}^{i}\right)=\sum_{j\in\mathcal{N}_{i}}q_{i}\left(t_{k}^{i}\right)-q_{j}\left(t_{k}^{i}\right)\!+\!g_i\left(q_{i}\left(t_{k}^{i}\right)-q_{r}\left(t_{k}^{i}\right)\right),
\end{equation*}
and $\varphi_i\left(t_{k}^{i}\right)=\alpha \xi_i\left(t_{k}^{i}\right)+\beta \zeta_i\left(t_{k}^{i}\right)$.
By denoting the measurement error of agent $i$ as
$
e_{i}(t)=u_i(t_k^i)-u_i(t).
$
The triggering instant sequence $t_{k}^{i}$ for agent $i$ is defined iteratively by
\begin{equation}
t_{k+1}^{i}=\inf \left\{t>t_{k}^{i}: f_{i}(t) \geq 0\right\},
\end{equation}
where the event-triggering function $f_i(t)$ corresponding to agent $i$ is
\begin{equation}\label{eq:15.5}
f_i(t)=\lVert e_i(t)\rVert-\rho\lVert \varphi_i(t)\rVert,
\end{equation}
where $\rho >0$. Then, according to the proposed control protocol (\ref{eq:14}), the dynamics of agent $i$ can be written into the following compact form
\begin{equation}\label{eq:15.75}
\left\{\begin{array}{l}
{\dot{\underline{s}}(t)=\underline{q}(t)} \\
{\dot{\underline{q}}(t)=\underline{e}(t)\!-\!k_1\operatorname{sig}\left(\underline{\varphi}\left(t\right)\right)^{\gamma_{1}}\!-\!k_2\operatorname{sig}\left(\underline{\varphi}\left(t\right)\right)^{\gamma_{2}}\!- \! k_3\underline{\varphi}\left(t\right)}
\end{array}\right.,
\end{equation}
where $t \in\left[t_{k}^{i}, t_{k+1}^{i}\right)$. By taking the transformations $\hat{s}_i(t)=s_i(t)-s_r(t)$, $\hat{q}_i(t)=s_i(t)-s_r(t)$, we obtain $\underline{\xi}(t)=(L+G)\underline{\hat{s}}(t)$ and $\underline{\zeta}(t)=(L+G)\underline{\hat{q}}(t)$. Hence, from (\ref{eq:15.75}), one has
\begin{equation}\label{eq:16}
\left\{\begin{array}{l}
{\dot{\underline{\xi}}(t)=\underline{\zeta}(t)} \\
{\dot{\underline{\zeta}}(t)\!=}\\
{Q\left\{\underline{e}(t)\!-\!k_1\operatorname{sig}\left(\underline{\varphi}\left(t\right)\right)^{\gamma_{1}}\!-\!k_2\operatorname{sig}\left(\underline{\varphi}\left(t\right)\right)^{\gamma_{2}}\!-\! k_3\underline{\varphi}\left(t\right)\right\}}
\end{array}\right.,
\end{equation}
where $Q=L+G$. Let $ \underline{\varepsilon}(t)=\left(\underline{\xi}^T(t), \underline{\zeta}^T(t)\right)^T$, from (\ref{eq:16}), we can get that
\begin{equation}\label{eq:16.5}
\dot{\underline{\varepsilon}}=M\underline{\varepsilon}+N,
\end{equation}
where $M=\left[\begin{smallmatrix}{\boldsymbol{O}_{|\mathcal{B}|}} & {\boldsymbol{I}_{|\mathcal{B}|}} \\ {-\alpha Q} & {-\beta Q}\end{smallmatrix}\right]\in \mathbb{R}^{2|\mathcal{B}|\times 2|\mathcal{B}|}$, $N=\left[\begin{smallmatrix}{\boldsymbol{0}_{|\mathcal{B}|}}  \\ {Q\left\{\underline{e}(t)\!-\!k_1\operatorname{sig}\left(\underline{\varphi}\left(t\right)\right)^{\gamma_{1}}\!-\!k_2\operatorname{sig}\left(\underline{\varphi}\left(t\right)\right)^{\gamma_{2}} \!-\!(k_3-1)\underline{\varphi}(t)\right\}} \end{smallmatrix}\right]\in \mathbb{R}^{2|\mathcal{B}|\times 1}$. Based on the control protocol (\ref{eq:14}), we are now in a position
to present some criteria to guarantee the consensus of the studied MAS (\ref{eq:16.5}). We will use $\underline{\hat{s}}$, $\underline{\hat{q}}$, $\underline{\xi}$, $\underline{\zeta}$, $\underline{\varphi}$, $\underline{\varepsilon}$ and $\underline{e}$ to represent $\underline{\hat{s}}(t)$, $\underline{\hat{q}(t)}$, $\underline{\varphi}(t)$, $\underline{\varphi}(t)$, $\underline{\varphi}(t)$, $\underline{\varepsilon}(t)$ and $\underline{e}(t)$ for brevity. Before giving the main result, some supporting lemmas are provided as follows

\textbf{\emph{Lemma 1:}}\label{lem:1} \cite{171} Consider the system $\dot{\underline{x}}=f(\underline{x})$ where $\underline{x} \in \mathbb{R}^n$, $f:\mathbb{R}^n \to \mathbb{R}^n$ is continuous and $f(\boldsymbol{0}_n)=\boldsymbol{0}_n$, if $f(\underline{x})$ is a homogenous vector function in the $0$-limit and $\infty$-limit with the associated triples $(\underline{r}_\mathrm{z},d_{\mathrm{z}},f_{\mathrm{z}})$ and $(\underline{r}_{\mathrm{I}},d_{\mathrm{I}},f_{\mathrm{I}})$, respectively. Moreover, the original system $\dot{\underline{x}}=f(\underline{x})$ and the approximating systems $\dot{\underline{x}}=f_{\mathrm{z}}(\underline{x})$ and $\dot{\underline{x}}=f_{\mathrm{I}}(\underline{x})$ are globally asymptotically stable. Then, the following condition is satisfied: Let $d_{V_{\mathrm{z}}}$ and $d_{V_{\mathrm{I}}}$ be real numbers such that $d_{V_{\mathrm{z}}}>\max_{1\leq i \leq n} r_{\mathrm{z},i}$ and $d_{V_{\mathrm{I}}}>\max_{1\leq i \leq n} r_{\mathrm{I},i}$. Then, there exists a positive definite function $V:\mathbb{R}^n\to \mathbb{R}_+$, such that, for $i\in \{1,\cdots,n\}$,
\begin{equation}
\frac{\partial V}{\partial x}(x) f(x) \leq-c_{V} \mathfrak{H}\left(V(x)^{\frac{d_{V_{\mathrm{z}}}+d_{\mathrm{z}}}{d_{V_{\mathrm{z}}}}}, V(x)^{\frac{d_{V_{\mathrm{I}}+d_{\mathrm{I}}}}{d_{V_{\mathrm{I}}}}}\right),
\end{equation}
where $c_V$ is positive real constant.

\textbf{\emph{Lemma 2:}}\label{lem:2} \cite{181} If there have a continuous radially unbounded function $V:\mathbb{R}_+^2 \to \mathbb{R}_+$ such that 1) $V(\underline{x})=0 \Leftrightarrow \underline{x}=\boldsymbol{0}_n$, 2) the solution $\underline{x}(t)$ satisfies the inequality $D^{*} V(\underline{x}(t)) \leq -\left(\tilde{\alpha} V^{\tilde{p}}(x(t))+\tilde{\beta} V^{\tilde{q}}(x(t))\right)^{r}$ for some $\tilde{\alpha}, \tilde{\beta}, \tilde{p}, \tilde{q}, r>0$ : $ \tilde{p} r<1, \tilde{q} r>1$, then $\underline{x}=\boldsymbol{0}_n$ is globally fixed-time attractive and the settling
time $T_f \leq \frac{1}{\tilde{\alpha}^{r}(1-\tilde{p} r)}+\frac{1}{\tilde{\beta}^{r}(\tilde{q} r-1)}$.

\textbf{\emph{Theorem 1:}}\label{thm:1}
\rm Using control protocol (\ref{eq:14}), the frequency restoration and SoC balancing problem in (\ref{eq:12.5}) is solved in a finite time that is irrelevant to initial conditions, if the event-triggering function (\ref{eq:15.5}) is enforced to be no greater that zero and the following conditions are satisfied
\begin{equation}\label{con:1}2\beta^2-\alpha \lambda_{|\mathcal{B}|}(Q^{-1})>0 ,\end{equation}
\begin{equation}\label{con:2}k_3-1-d / 2+\alpha^2-\beta^2+ \lambda_{|\mathcal{B}|}(Q^{-1})>0,\end{equation}
\begin{equation}\label{con:3}0<\rho<2d(k_3-1-d / 2+\alpha^2-\beta^2+ \lambda_{|\mathcal{B}|}(Q^{-1})),\end{equation}
where $d>0$. Moreover, the upper bound of the settling time is estimated as
\begin{equation}
T_{f} \leq \frac{2}{c_{V}\left(1-{\frac{d_{V_{\mathrm{z}}}+d_{\mathrm{z}}}{d_{V_{\mathrm{z}}}}}\right)}+\frac{2}{c_{V}\left({\frac{d_{V_{\mathrm{I}}+d_{\mathrm{I}}}}{d_{V_{\mathrm{I}}}}}-1\right)},
\end{equation}
where constants $c_{V}>0$, $d_{V_{\mathrm{z}}},d_{V_{\mathrm{I}}} >0$, $d_{\mathrm{z}}<0$ and $d_{\mathrm{I}}>0$.
\noindent
\quad \emph{Proof}: The proof will be divided into two steps. Firstly, the global asymptotical stability of the closed-loop system (\ref{eq:15.75}) is established. Next, we prove that the origin of system (\ref{eq:15.75}) under protocol (\ref{eq:14}) with $t \in\left[t_{k}^{i}, t_{k+1}^{i}\right)$ is finite-time stable.
\par \emph{Step 1}: Consider the Lyapunov candidate function
$
V_1(t)=\frac{1}{2}\underline{\varepsilon}^TP\underline{\varepsilon},
$
with $P=\left[\begin{smallmatrix}{2\alpha\beta\boldsymbol{I}_{|\mathcal{B}|}} & {\alpha Q^{-1}} \\ {\alpha Q^{-1}} & {\beta Q^{-1}}\end{smallmatrix}\right]\in \mathbb{R}^{2|\mathcal{B}|\times 2|\mathcal{B}|}$. Since $Q^{-1}$ is positive definite, using Schur complements, $P$ is positive definite if and only if
\begin{equation}
2\alpha\beta\boldsymbol{I}_{|\mathcal{B}|}-\left(\alpha Q^{-1}\right)\left(\frac{1}{\beta}Q\right)\left(\alpha Q^{-1}\right)>0.
\end{equation}
which implies that there exits an invertible matrix $\Gamma\in \mathbb{R}^{|\mathcal{B}|\times |\mathcal{B}|}$, such that $\Gamma^{-1}\left( 2\beta^2\boldsymbol{I}_{|\mathcal{B}|}-\alpha \operatorname{diag}\left(\{ \lambda_i(Q^{-1})\} \right)\right)\Gamma>0$. Then, One have that if $2\beta^2-\alpha \lambda_{i}(Q^{-1})>0$ for all $i\in \mathcal{B}$, $\frac{1}{2}\underline{\varepsilon}^TP\underline{\varepsilon}>0$. From condition (\ref{con:1}), $V_1(t)$ is positive definite and it equals to zero if and only if the consensus is reached.
\par Differentiating $V_1(t)$ yields
\begin{equation}\label{eq:18}
\begin{aligned}
\dot{V}_1(t)=&\underline{\varepsilon}^TP\dot{\underline{\varepsilon}}
=\underline{\varepsilon}^TP\left( M\underline{\varepsilon}+N\right)\\
=&-\alpha^2\underline{\varphi}^T\underline{\varphi}-\underline{\varphi}^T(\beta^2\boldsymbol{I}_{|\mathcal{B}|}-\alpha Q^{-1})\underline{\varphi}+\underline{\varphi}^T\underline{e}
\\&-k_1\underline{\varphi}^T\operatorname{sig}\left(\underline{\varphi}\right)^{\gamma_{1}}- k_2\underline{\varphi}^T\operatorname{sig}\left(\underline{\varphi}\right)^{\gamma_{2}}-(k_3-1)\underline{\varphi}^T\underline{\varphi}.
\end{aligned}
\end{equation}
By use of Young’s inequality, we obtain
\begin{equation}\label{eq:19}
\underline{\varphi}^T\underline{e}
\leq(d / 2) \underline{\varphi}^{T} \underline{\varphi}+(1 / 2 d) \underline{e}^{T}\underline{e},
\end{equation}
for all $d>0$. Furthermore, it follows from the Courant-Fischer Theorem that
\begin{equation}\label{eq:20}
\begin{aligned}
&\underline{\varphi}^T(\beta^2\boldsymbol{I}_{|\mathcal{B}|}-\alpha Q^{-1})\underline{\varphi}\\&
=\underline{\varphi}^T\Xi^{-1}\left( \beta^2\boldsymbol{I}_{|\mathcal{B}|}-\operatorname{diag}\left(\{ \lambda_i(Q^{-1})\} \right)\right)\Xi\underline{\varphi}\\
&\geq \lambda_1\left\{\Xi^{-1}\left( \beta^2\boldsymbol{I}_{|\mathcal{B}|}-\operatorname{diag}\left(\{ \lambda_i(Q^{-1})\} \right)\right)\Xi\right\}\underline{\varphi}^T\underline{\varphi}\\
&\geq \lambda_1\left\{\Xi^{-1}\left( \beta^2\boldsymbol{I}_{|\mathcal{B}|}-\operatorname{diag}\left(\{ \lambda_{|\mathcal{B}|}(Q^{-1})\} \right)\right)\Xi\right\}\underline{\varphi}^T\underline{\varphi}\\
&= \left(\beta^2- \lambda_{|\mathcal{B}|}(Q^{-1})\right)\underline{\varphi}^T\underline{\varphi}.
\end{aligned}
\end{equation}
Combining (\ref{eq:18}), (\ref{eq:19}) and (\ref{eq:20}), we obtain
\begin{equation}\label{eq:21}
\begin{aligned}
\dot{V}_1(t)\!&\leq\! -k_1\underline{\varphi}^T\operatorname{sig}\left(\underline{\varphi}\right)^{\gamma_{1}}\!- \! k_2\underline{\varphi}^T\operatorname{sig}\left(\underline{\varphi}\right)^{\gamma_{2}}\!-\\
&(k_3-1-d / 2+\alpha^2\!-\!\beta^2\!+\! \lambda_{|\mathcal{B}|}(Q^{-1})) \underline{\varphi}^{T} \underline{\varphi}+(1 / 2 d) \underline{e}^{T}\underline{e}.
\end{aligned}
\end{equation}
From conditions (\ref{con:2}) and (\ref{con:3}), if the event-triggering function (\ref{eq:15.5}) can be enforced to be no greater than zero, one then has
\begin{equation}\label{eq:23}
\begin{aligned}
\dot{V}_1(t)&\leq -\underline{\varphi}^T\operatorname{sig}\left(\underline{\varphi}\right)^{\gamma_{1}}- \underline{\varphi}^T\operatorname{sig}\left(\underline{\varphi}\right)^{\gamma_{2}}\\
&=-\sum_{i\in\mathcal{B}}|\varphi_i|^{\gamma_{1}+1}-\sum_{i\in\mathcal{B}}|\varphi_i|^{\gamma_{2}+1}\leq 0.
\end{aligned}
\end{equation}
Then it follows from (\ref{eq:23}) that the closed-loop system (\ref{eq:15.75}) is global asymptotically stable.
\par \emph{Step 2}: The the closed-loop system (\ref{eq:15.75}) can be rewritten as follows
\begin{equation}\label{eq:24}
\left\{\begin{array}{l}
{\dot{\hat{s}}_i=\hat{q}_i} \\
{\dot{\hat{q}}_i=-k_1\operatorname{sig}\left(\varphi_i\right)^{\gamma_{1}}+\mathfrak{g}(\hat{s}_i,\hat{q}_i)}
\end{array}\right., i\in \mathcal{B},
\end{equation}
where
\begin{equation}\label{eq:24.5}
\begin{aligned}
&\mathfrak{g}(\hat{s}_i,\hat{q}_i)=-k_2\operatorname{sig}\left(\varphi_i\right)^{\gamma_{2}}-k_3\varphi_i+e_i.
\end{aligned}
\end{equation}
Next, we consider the  candidate approximating system of system (\ref{eq:24.5})
\begin{equation}\label{eq:25}
\left\{\begin{array}{l}
{\dot{\hat{s}}_i=\hat{q}_i} \\
{\dot{\hat{q}}_i\!=-k_1\operatorname{sig}\left(\varphi_i\right)^{\gamma_{1}}}
\end{array}\right., i\in \mathcal{B},
\end{equation}
which is denoted as $\mathfrak{f}(\hat{s}_i,\hat{q}_i)$.
Noting that $0<\gamma_1<1$, it is easy to verify that
\begin{equation}\label{eq:26}
\left\{\begin{array}{l}
{\frac{\mathfrak{f}(\kappa^{r_{\mathrm{z},1}}\hat{s}_i,\kappa^{r_{\mathrm{z},2}}\hat{q}_i)}{\kappa^{r_{\mathrm{z},1}+d_{\mathrm{z}}}}=\mathfrak{f}(\hat{s}_i,\hat{q}_i)} \\
{\frac{\mathfrak{f}(\kappa^{r_{\mathrm{z},1}}\hat{s}_i,\kappa^{r_{\mathrm{z},2}}\hat{q}_i)}{\kappa^{r_{\mathrm{z},2}+d_{\mathrm{z}}}}=\mathfrak{f}(\hat{s}_i,\hat{q}_i)}
\end{array}\right.,
\end{equation}
where $\underline{r}_\mathrm{z}=[\frac{1}{1-\gamma_1},\frac{1}{1-\gamma_1}]$ and $d_{\mathrm{z}}=-1$. Then,  system (\ref{eq:25}) is homogenous of degree $d_{\mathrm{z}}=-1$ with respect to $\underline{r}_\mathrm{z}$. Furthermore, one can easily have
\begin{equation}\label{eq:27}
\lim _{\kappa \rightarrow 0} \frac{\mathfrak{g}\left(\kappa^{r_{\mathrm{z},1}} \hat{s}_{i}, \kappa^{r_{\mathrm{z},2}} \hat{q}_i\right)}{\kappa^{r_{\mathrm{z},2}+d_{\mathrm{z}}}}=0.
\end{equation}
Then, we can conclude that system (\ref{eq:25}) is the approximating systems of system (\ref{eq:24}) in the $0$-limit associated with triple $(\underline{r}_\mathrm{z},d_{\mathrm{z}},\mathfrak{f}(\hat{s}_i,\hat{q}_i))$.
Further, to establish the globally asymptotic stability of (\ref{eq:25}), we choose the Lyapunov candidate function as follows
\begin{equation}\label{eq:28}
V_2=\frac{\alpha}{2k_1}\underline{\hat{q}}^TQ\underline{\hat{q}}+\sum_{i \in \mathcal{B}}\int_{0}^{\varphi_i} k_1\left(\operatorname{sig}(s)^{\gamma_{1}}\right) d s.
\end{equation}
Note that $\varphi_i$ and $\operatorname{sig}(\varphi_i)^{\gamma_{1}}$ have same sign in sense of component-wise, then we have $\int_{0}^{\varphi_i} \left(\operatorname{sig}(s)^{\gamma_{1}}\right) d s>0$ for any $\varphi_i \neq 0$. Moreover, in the light of $Q>0$, we can conclude that $V_2$ is positive before the consensus is reached. Differentiating (\ref{eq:28}), we have
\begin{equation}\label{eq:29}
\begin{aligned}
\dot{V}_2&=\frac{\alpha}{k_1} \left( Q\underline{\hat{q}}\right)^T \dot{\underline{\hat{q}}}+\sum_{i \in \mathcal{B}}k_1 \dot{\varphi}_i\operatorname{sig}(\varphi_i)^{\gamma_{1}}\\
&=-\alpha \underline{\zeta}^T\operatorname{sig}\left(\underline{\varphi}\left(t\right)\right)^{\gamma_{1}}+k_1\dot{\underline{\varphi}}
\operatorname{sig}\left(\underline{\varphi}\left(t\right)\right)^{\gamma_{1}}\\
&=-\alpha \underline{\zeta}^T\operatorname{sig}\left(\underline{\varphi}\left(t\right)\right)^{\gamma_{1}}+\left(\alpha \underline{\zeta}+ \beta \dot{\underline{\zeta}}\right)^T\operatorname{sig}\left(\underline{\varphi}\left(t\right)\right)^{\gamma_{1}}\\
&= \left(\beta Q\dot{\underline{\hat{q}}}\right)^T\operatorname{sig}\left(\underline{\varphi}\left(t\right)\right)^{\gamma_{1}}\\
&=-\beta \left[\operatorname{sig}\left(\underline{\varphi}\left(t\right)\right)^{\gamma_{1}}\right]^TQ\operatorname{sig}\left(\underline{\varphi}\left(t\right)\right)^{\gamma_{1}}\leq 0,
\end{aligned}
\end{equation}
which implies that the origin of system (\ref{eq:25}) is globally asymptotically stable.

Rewrite the closed-loop system (\ref{eq:15.75}) as follows
\begin{equation}\label{eq:30}
\left\{\begin{array}{l}
{\dot{\hat{s}}_i=\hat{q}_i} \\
{\dot{\hat{q}}_i=-k_2\operatorname{sig}\left(\varphi_i\right)^{\gamma_{2}}+\tilde{\mathfrak{g}}(\xi_i,\zeta_i)}
\end{array}\right., i\in \mathcal{B},
\end{equation}
where $
\tilde{\mathfrak{g}}(\hat{s}_i,\hat{q}_i)=-k_2\operatorname{sig}\left(\varphi_i\right)^{\gamma_{2}}-k_3\varphi_i+e_i.
$
Next, we consider the candidate approximating system of system (\ref{eq:24.5})
\begin{equation}\label{eq:32}
\left\{\begin{array}{l}
{\dot{\hat{s}}_i=\hat{q}_i} \\
{\dot{\hat{q}}_i\!=-k_2\operatorname{sig}\left(\varphi_i\right)^{\gamma_{2}}}
\end{array}\right., i\in \mathcal{B},
\end{equation}
which is denoted as $\tilde{\mathfrak{f}}(\hat{s}_i,\hat{q}_i)$.
Similar to the analysis in \emph{Step 2}, it can be verified that system (\ref{eq:30}) is a homogenous vector function in the $\infty$-limit with the associated triple $(\underline{r}_{\mathrm{I}},d_{\mathrm{I}},\tilde{\mathfrak{f}}(\hat{s}_i,\hat{q}_i))$, where $\underline{r}_{\mathrm{I}}=[\frac{1}{\gamma_2-1},\frac{1}{\gamma_2-1}]$ and $d_{\mathrm{I}}=1$. Moreover, to establish the globally asymptotic stability of (\ref{eq:32}),
we choose the Lyapunov candidate function as follows
\begin{equation}\label{eq:33}
V_3=\frac{\alpha}{2k_2}\underline{\hat{q}}^TQ\underline{\hat{q}}+\sum_{i \in \mathcal{B}}\int_{0}^{\varphi_i} k_2\left(\operatorname{sig}(s)^{\gamma_{1}}\right) d s.
\end{equation}
Analogous with the proof in (\ref{eq:29}), it can be proved that the origin of system (\ref{eq:32}) is globally
asymptotically stable.

Thus, based on \emph{Step 1-2}, all conditions in \emph{Lemma 1} are satisfied, and there exists a positive constant $c_V$ and a positive definite  Lyapunov candidate function $V$ for the original system (\ref{eq:15.75}), such that
\begin{equation}\label{eq:34}
\dot{V} \leq-c_{V} \mathfrak{H}\left(V^{\frac{d_{V_{\mathrm{z}}}+d_{\mathrm{z}}}{d_{V_{\mathrm{z}}}}}, V^{\frac{d_{V_{\mathrm{I}}+d_{\mathrm{I}}}}{d_{V_{\mathrm{I}}}}}\right),
\end{equation}
where $d_{V_{\mathrm{z}}}=\max_{1\leq i \leq 2} r_{\mathrm{z},i}$ and $d_{V_{\mathrm{I}}}=\max_{1\leq i \leq 2} r_{\mathrm{I},i}$. Further, note that $0<\gamma_{1}<1$ and $\gamma_{2}>1$, we can drive that
\begin{equation}
\mathfrak{H}\left(V^{\frac{d_{V_{\mathrm{z}}}+d_{\mathrm{z}}}{d_{V_{\mathrm{z}}}}}, V^{\frac{d_{V_{\mathrm{I}}+d_{\mathrm{I}}}}{d_{V_{\mathrm{I}}}}}\right) \geq \frac{1}{2}\left(V^{\frac{d_{V_{\mathrm{z}}}+d_{\mathrm{z}}}{d_{V_{\mathrm{z}}}}}+ V^{\frac{d_{V_{\mathrm{I}}+d_{\mathrm{I}}}}{d_{V_{\mathrm{I}}}}}\right),
\end{equation}
which follows that
\begin{equation}
\dot{V} \leq \frac{-c_{V}}{2} \left(V^{\frac{d_{V_{\mathrm{z}}}+d_{\mathrm{z}}}{d_{V_{\mathrm{z}}}}}+ V^{\frac{d_{V_{\mathrm{I}}+d_{\mathrm{I}}}}{d_{V_{\mathrm{I}}}}}\right).
\end{equation}
By \cite{181}, the consensus of the MAS is achieved within a settling time that satisfied
\begin{equation}
T_{f} \leq \frac{2}{c_{V}\left(1-{\frac{d_{V_{\mathrm{z}}}+d_{\mathrm{z}}}{d_{V_{\mathrm{z}}}}}\right)}+\frac{2}{c_{V}\left({\frac{d_{V_{\mathrm{I}}+d_{\mathrm{I}}}}{d_{V_{\mathrm{I}}}}}-1\right)}.
\end{equation}
This together with \emph{Lemma 6} in \cite{22} give that
the origin of system (\ref{eq:15.75}) is globally finite-time stable. The proof is completed.\hfill $\blacksquare$

\textbf{\emph{Remark 1:}}
In some work \cite{5,6}, the communication in the secondary control is considered directed, and the neighbor of agent $i$, agent $j$, is triggered at time $t^j_{k'}$, where $k^{\prime}(t) \triangleq \arg \min _{l \in \mathbb{N}: t \geq t_{l}^{j}}\{t-t_{l}^{j}\}$. It is known that $t^j_{k'}$ may not be equal to $t^i_{k}$. In this case, the control input will update at triggering instants of both agent $i$ and $j$, which may cause more frequent triggered events. However, in our work, the communication network is undirected such that a bidirectional communication is feasible. Then, if agent $i$ is triggered at $t^i_{k}$, one channel is first set up to send time $t^i_{k}$ to its neighbor, agent $j$, and another channel is utilized to transmit the signal from agent $j$ to agent $i$. The proposed design in this paper apply to most industrial communication protocols that are bidirectional, such as TCP, and may efficiently reduce triggered events.

{\color{blue}\textbf{\emph{Remark 2:}}
In the proposed control algorithm (\ref{eq:14}), the control gains $k_1>0$, $k_2>0$, $k_3>0$, $\alpha>0$ and $\beta>0$ can be determined in an online adaptive or an off-line method, which can be determined by considering a trade-off among convergence time, overshoot for oscillation of dynamic process, event interval and sensitivity to the measured noise. 
\begin{table}[htbp]
\renewcommand\arraystretch{1.3}
\centering
\caption{Control performance with different control parameters}
\begin{tabular}{p{3cm}p{2.8cm}<{\centering}p{2.8cm}<{\centering}p{2.8cm}<{\centering}}
\hline
  \toprule
Gains $k_1,k_2,k_3$ and $\alpha,\beta$ & $(k_1,k_2,k_3,\alpha,\beta)\in S_1$ & $(k_1,k_2,k_3,\alpha,\beta)\in S_2$ & $(k_1,k_2,k_3,\alpha,\beta)\in S_3$ \\
  \toprule
Conv. Time $(q_i)$ &4.89s &2.32s&2.17s \\

Oversh. $(q_i)$ & 10.9\% &11.5\%&25.8\%\\
Event Int. &1.64ms & 13.1ms& 10.5ms\\

  \bottomrule
\end{tabular}
\end{table}
%%%%%%%%%%%%%%%%%%%%%%%%%%%%%%%%%%%%%%%%%%%%%%%%%  Table II
To further characterize the selection method of control gains, we also simulate the proposed controller with three groups of $(k_i,c_i)$, $S_1=\{30,35,16,72,12\}$, $S_2=\{6,7,3.5,72,12\}$, and $S_3=\{6,7,3.5,150,20\}$, the convergence time, transient overshoot and the average event interval are given in Table I. It indicates that a fast convergence speed can be obtained by increasing gains at the sacrifice of large overshoots and event interval. Based on the off-line calculation, we conclude that the setting of $k_1,k_2,k_3$ and $\alpha,\beta$ of $S_2$ in Table I is reasonable. Under this condition, it guarantees the system maintaining desired convergence time, certain transient overshoot and event interval. It further validates the flexibility in gain selection.}

\emph{B. Event Interval Analysis}

In the following, the inter-event times $t_{k+1}^{i}- t_{k}^{i}$, $i\in \mathcal{B}$, should be proven to have a lower bound for $k\in \mathbb{N}$ so that the Zeno behavior is able to be avoided. If the event-triggering function (\ref{eq:15.5}) of agent $i$ reaches zero at time $t_{k}^{i}+\tau_{i}$, then one has
\begin{equation}\label{35}
\begin{aligned}
&\left\|e_{i}\left(t_{k}^{i}+\tau_{i}\right)\right\|=\left\|\int_{t_{k}^{i}}^{t_{k}^{i}+\tau_{i}} \dot{e}_{i}(s) \mathrm{d} s\right\|=\\
&\left\|\int_{t_{k}^{i}}^{t_{k}^{i}+\tau_{i}} \dot{\varphi}_i \left(k_1\gamma_1\varphi_i^{\gamma_1-1}+k_2\gamma_1\varphi_i^{\gamma_2-1}+k_3\right)\mathrm{d} s\right\|\\
&\leq \dot{\varphi}_i\left\| k_1\gamma_1\varphi_i^{\gamma_1-1}+k_2\gamma_1\varphi_i^{\gamma_2-1}+k_3 \right \|\tau_{i}.
\end{aligned}
\end{equation}
Let $\phi_i=\max \dot{\varphi}_i\left\| k_1\gamma_1\varphi_i^{\gamma_1-1}+k_2\gamma_1\varphi_i^{\gamma_2-1}+k_3 \right \|$. Obviously, it is seen that $\phi_i>0$ before the consensus is achieved. Then, we have
\begin{equation}
\tau_{i} \geq {\left\|e_{i}\left(t_{k}^{i}+\tau_{i}\right)\right\|}/{\phi_i}.
\end{equation}
Since the next triggering instant $t_{k+1}^{i}$ of agent $i$ happens when the event-triggering function (\ref{eq:15.5}) is equal to or greater than zero, one can obtain $\left\|e_{i}\left(t_{k}^{i}+\tau_{i}\right)\right\| \geq 0$ at $t_{k+1}^{i}$.
Hence, the event interval $\tau_i=t_{k+1}^{i}-t_{k}^{i}$ satisfies the following inequality
\begin{equation}
\tau_{i} \geq {(k_3-1)\lVert \varphi_i(t)\rVert}/{\phi_i}>0,
\end{equation}
before the consensus is achieved, which ensures that the Zeno behavior is able to be avoided.

{\color{blue}\textbf{\emph{Remark 3:}}
The event-triggered control and finite-time control take effect at the same time. Under the event-triggered mechanism, the control input $u_i(t)$ is hold constant during the event intervals, and the trajectory of the state variable is no longer smooth. It is proved in Theorem 1 that the non-smooth system is finite-time stable. According to the above analysis, the proposed control can avoid Zeno behaviour if the stable state is not reached, that is, $t\leq T_f$. Since, after $t=T_f$, $\rho\lVert \varphi_i(t)\rVert$ strictly equals to zero, it follows that the measurement error $e_i(t)$ maintains zero, which means that the events would not be triggered.}

\emph{C. Improved Event-Triggered Strategy}

Based on the above analysis, the proposed event-triggering function (\ref{eq:15.5}) requires continuous information exchange between connected agents in the calculation of the event instants. Hence, in order to further reduce
the communication burdens and lessen calculative burden, an improved event-triggered strategy is developed. This improved event-triggered strategy only uses the discrete states that sampled and sent by neighbors at their event instants. Note that the term $\varphi_i(t)$ makes the event-triggering process require continuous communication, we thus need to reformulate $\varphi_i(t)$. According to (\ref{eq:16}), we have
\begin{equation}\label{eq:36}
\begin{aligned}
&\zeta_i(t)=\int_{t_{k}^{i}}^{t}\dot{\zeta}_i(s)ds+{\zeta}_i(t_{k}^{i})\\
&=\left(\int_{t_{0}^{\prime}}^{t_{1}^{\prime}}+\int_{t_{1}^{\prime}}^{t_{2}^{\prime}}+\ldots+\int_{t_{1}^{\prime}}^{t}\right)\dot{\zeta}_i(t)+{\zeta}_i(t_{k}^{i})\\
&=\sum_{h=0}^{l-1}\dot{\zeta}_i(t_{h}^{\prime})\left(t_{h+1}^{\prime}-t_{h}^{\prime}\right)+\dot{\zeta}_i(t_{l}^{\prime})\left(t-t_{l}^{\prime}\right)+\zeta_{i}\left(t_{k}^{i}\right),
\end{aligned}
\end{equation}
where $\left[t_{k}^{i}, t_{k+1}^{i}\right)$ and $t_{k}^{i}=t_{0}^{\prime}<t_{1}^{\prime}<\cdots<t_{l}^{\prime}<t_{l+1}^{\prime}=t_{k+1}^{i}$ are the updating instants of $\zeta_i(t)$. Further, from (\ref{eq:36}), $\xi_i(t)$ can be written as
\begin{equation}\label{eq:37}
\begin{aligned}
&\xi_i(t)=\\
&\left(\zeta_{i}\left(t_{k}^{i}\right)-\dot{\zeta}_i(t_{l}^{\prime}) t_{l}^{\prime}+\sum_{k=0}^{l-1} \dot{\zeta}_i(t_{h}^{\prime})\left(t_{h+1}^{\prime}-t_{h}^{\prime}\right)\right)\left(t-t_{k}^{i}\right)\\&+\frac{\dot{\zeta}_i(t_{l}^{\prime})}{2}\left(t^{2}-\left(t_{k}^{i}\right)^{2}\right)
+\xi_{i}\left(t_{k}^{i}\right).
\end{aligned}
\end{equation}
Since $\varphi_i(t)=\alpha \xi_i\left(t\right)+\beta \zeta_i\left(t\right)$, $\varphi_i(t)$ can be calculated by the discrete information, and the continuous information exchange is avoided.

\begin{figure}[!t]
\centering
\includegraphics[width=0.5\columnwidth]{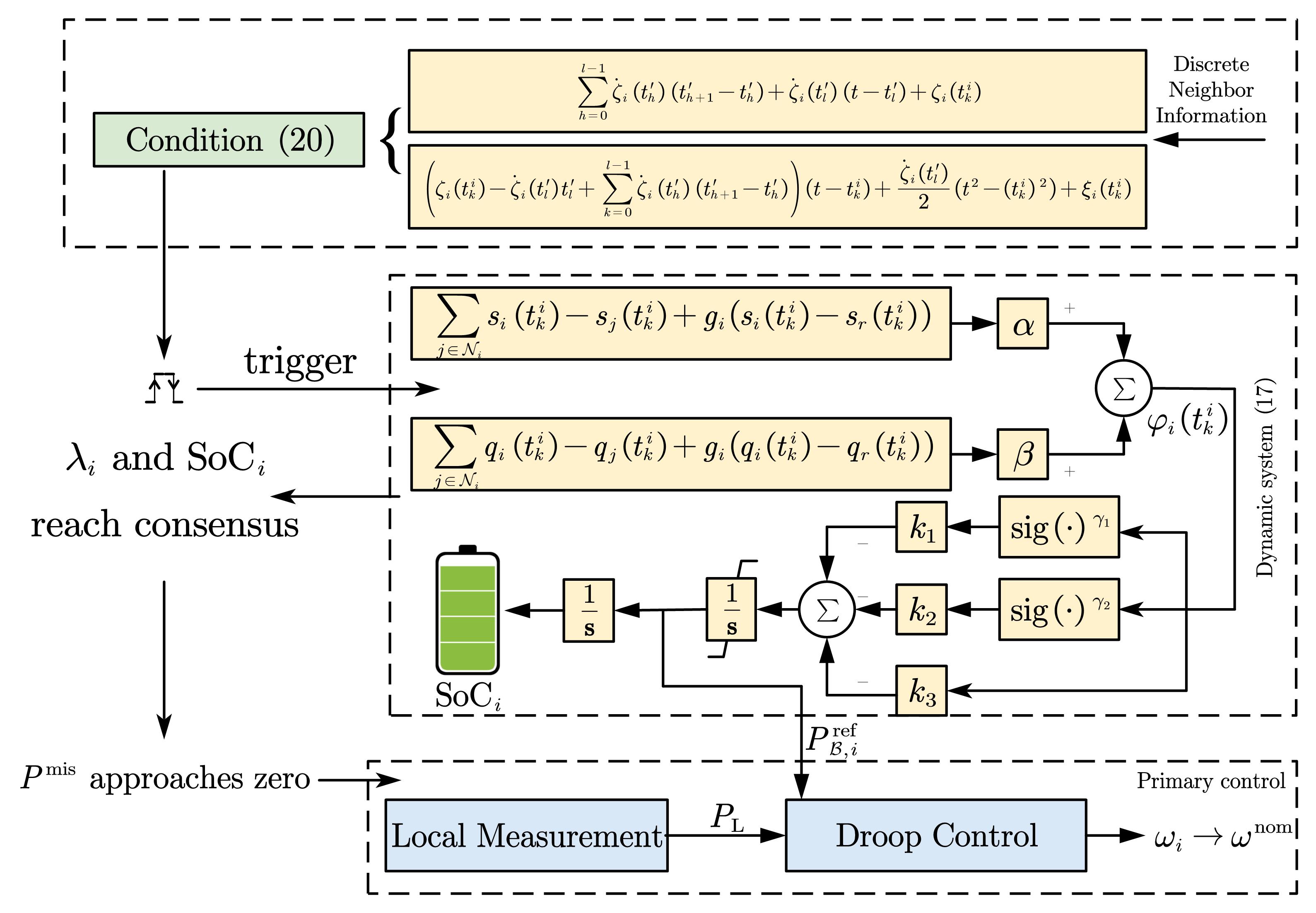}
\caption{Diagram of the proposed control scheme.}\label{fig:4}
\end{figure}
\begin{algorithm}[t!]
\caption{Implementation of the proposed control scheme}
\label{alg:Framwork}
\begin{algorithmic}[1] %这个1 表示每一行都显示数字
\REQUIRE ~~\\ %算法的输入参数：Input
1. Set $t_0^i=0$, $e_i(0)=0$ and $k=0$; Initializes the state information.\\
2. Control the local measurement module to track $P_\mathrm{L}$ on a time period $T$ cycle at leader nodes.\\
\ENSURE ~~\\ %算法的输出：Output
\IF{$\omega^\mathrm{syn}\in2 \pi \cdot [-0.1,0.1]$}
\STATE Hold $P_{\mathcal{B},i}^{\mathrm{ref}}$ in the previous state;
\ELSE
     \STATE Agent $i$ exchanges information with its neighbors and compute $\varphi_i(0)$.
     \STATE Compute $e_i(t)$ and $\varphi_i(t)$ according to (\ref{eq:15.5}), (\ref{eq:36}) and (\ref{eq:37});\\
     \IF{$f_i(t)\geq 0$}
     \STATE Set $t^i_k= t$, agent $i$ requires information $s_{j}\left(t_{k}^{i}\right)$ and $s_{j}\left(t_{k}^{i}\right)$ from its neighbors; Update $k = k + 1$, $\varphi_i(t^i_k)=\varphi_i(t)$ and reset $e_i(t ) = 0$;
     \ELSE
     \STATE Hold the input $u_i(t)$;
     \ENDIF
     \STATE $\lambda_{i}(t)=\lambda^*(t)$ and $\mathrm{SoC}_i(t)=\mathrm{SoC}_j(t)\,\forall i,j \in \mathcal{B}$;
\ENDIF
\IF{$s_{i}(t)\geq 80\%$ $||$ $s_{i}(t)\leq 20\%$}
     \STATE Set $q_{i}(t)=0$;
     \ELSE
     \STATE Continue;
     \ENDIF
\STATE Compute $P_{\mathcal{B},i}^{\mathrm{ref}}$ signal and input into the corresponding BESS. Then go back to 1.
\end{algorithmic}
\end{algorithm}

\emph{D. Control Algorithm Implementation}

{\color{blue}The proposed distributed controller shown in Fig. \ref{fig:4} is based on a multi-agent system, in which all agents negotiate an agreement using a consensus-based control protocol (\ref{eq:14}). The detailed implementation process is given in Algorithm 1. As seen, the controller in each agent deploys a primary and secondary control strategy. The local measurement at leader nodes measures the total power mismatch locally. During the iterative, primary control is first implemented, which requires only local information. Then utilizing the distributed control protocol (\ref{eq:14}), the secondary controllers coordinate $\lambda_i(t)$ to reach a consensus under the proposed event-triggered mechanism that is implemented by triggering condition (20) with discrete information. Additionally, it provides the reference charging/discharging power $P_{\mathcal{B}, i}^{\mathrm{ref}}$ for BESSs such that the total power mismatch $P^\mathrm{mis}$ is filled up. Then, the integrator, corresponding to state variables $\lambda_i(t)$, is set as a limited integrator with range $[-1, 1]$ to prevent the output from exceeding charging/discharging power bounds in the transient process. By (8), the designed secondary and the primary controllers thereby continuously drive the nominal active power terms to the output active power for all inverters as well as enable the frequencies to converge to the nominal value. Further, types of local constraints (SoC constraint and frequency deviation constraint) are integrated in the proposed distributed algorithm to prolong the battery life span. Firstly, the SoC of a battery is controlled within a proper range to avoid over-charging/over-discharging. Secondly, the BESS will be controlled to remain in the previous mode by including another projection operation if the frequency deviation of the system is acceptable (within 2\%), which may avoid a frequent switching between charging/discharging states.}
\begin{figure}[!t]
	\centering
	\includegraphics[width=0.5\columnwidth]{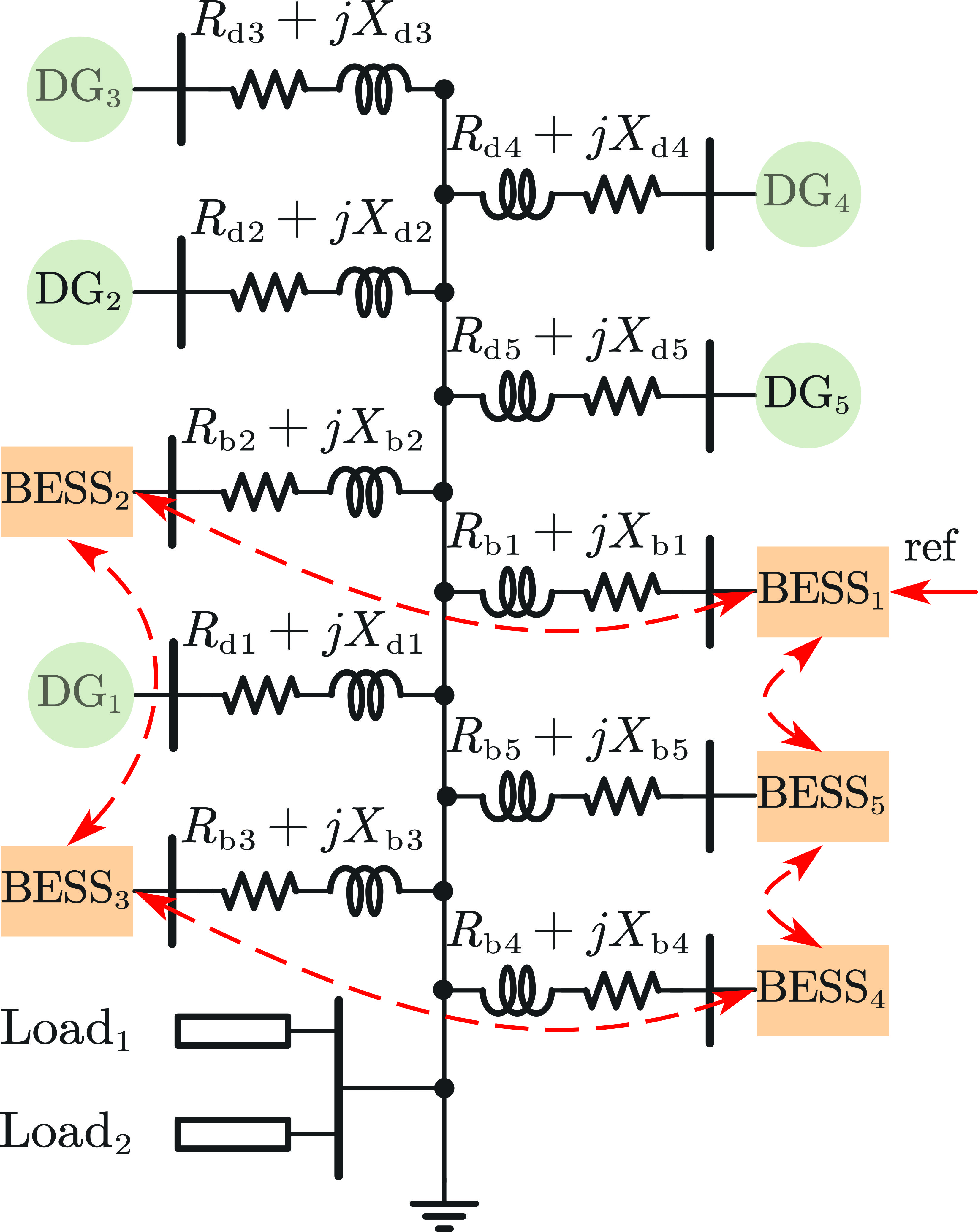}
	\caption{ AC microgrid test setup. The red dotted lines between BESSs represent communication links.}\label{fig:sim}
\end{figure}
\begin{table}[!t]
\renewcommand{\arraystretch}{1.3}
 \caption{Parameter values for simulation.}\label{tab:2}
 \centering
 \begin{tabular}{p{58pt}>{\raggedleft\arraybackslash}p{292pt}}
  \toprule
 Parameter& Value \\
  \midrule
 Nom. Freq & $\omega ^\mathrm{nom}/2\pi$=$50$Hz \\
 Nom. Volt&$V^{\mathrm{nom}}=311$V\\
 DGs:&$K_i^\mathrm{P}  \in\left\{ 0.02,0.02,0.02,0.02,0.02,0.02\right\} \text{MW}\text{s}$\\
 &$P_{\mathcal{R},i}^{\mathrm{nom}}\in \left\{5.0,8.0,15,12,10\right\} \text{kW}$\\
BESSs:&$K_i^\mathrm{P}  \in\left\{ 0.02,0.02,0.02,0.02,0.02,0.02\right\} -1\cdot\text{MW}\text{s}$\\
 &$P_{\mathcal{R},i}^{\mathrm{dis}}\in \left\{25,35,25,32,28\right\} \text{kW}$\\
& $P_{\mathcal{B},i}^\mathrm{cha}\in \left\{-25,-35,-25,-32,-28\right\}\text{kW}$\\
 Line. Impe:& $R_{\mathrm{d}i}\in\left\{0.23,0.3,0.12,0.15,0.25 \right\}\Omega$
$X_{\mathrm{d}i}\in\left\{3.0,2.9,3.1,3.0,3.0,3.0 \right\}\text{mH}$\\
& $R_{\mathrm{b}i}\in\left\{0.13,0.15,0.12,0.15,0.14 \right\}\Omega$
$X_{\mathrm{b}i}\in\left\{3.1,3.0,3.1,3.0,3.0,3.0 \right\}\text{mH}$\\
Cont. Para:& Case 1-2: $\alpha=72$, $\beta=12$, $k_1=6$, $k_2=7$, $k_3=3.5$\\
& Case 3: $\alpha=18$, $\beta=6$, $k_1=2$, $k_2=3$, $k_3=3.5$\\
  \bottomrule
 \end{tabular}
\end{table}
\section{Simulation}
In this section, the effectiveness of the proposed control scheme is verified by simulating multiple islanded ac microgrid cases, using MATLAB/SIMULINK. As depicted in Fig. \ref{fig:sim}, the test system contains five DGs and five BESSs, all of which are in parallel and communicate with each other through a undirected ring network. The simulation parameters are given in Table \ref{tab:2}. The following results cover following scenarios: 1) Without consideration for local constraints introduced above, Case 1 examine the restoration of the microgrid frequency and the balance of BESS SoC with the proposed improved distributed finite-time control algorithm (\ref{eq:14}) by inducing load/power changing; at the same time, the event intervals are recorded to verify anti-Zeno characteristic of the designed event-triggered mechanism. 2) Case 2 studies the performance of the proposed controller with the consideration of local constraints. 3) A comparative analysis is conducted among the proposed method, the finite-time controller in \cite{12,13} and the conventional asymptotic controller in \cite{191}.

\emph{Case 1: Frequency Restoration and SoC Balance}

This simulation study is divided into five stages. The first stage (0-3s) is the startup stage, in which the total load varies from $0$ to $70$kW, and the nominal power of DGs 1-5 are $5.0$, $8.0$, $15$, $12$ and $10$ respectively. The total power mismatch in this stage are then $-20$kW. Since the controller is not activated in this stage, the SoC of BESSs remains unchanged, $\lambda_i(t)$ of BESSs are all kept zero, and the steady-state frequency of the system deviates from $50$Hz, as depicted in Figs. 6c, d and e; At the beginning of the second stage (3-6s), the controller is activated. Next, all the $\lambda_i(t)$ automatically reach the consensus $\lambda^*=0.28$ within a settling time after a short oscillation, and the overshoots are tightly kept within the operational limits, as shown in Fig. 6b. It can be seen in Fig. 6c that the SoC balance is achieved increase as $\lambda_i(t)$ converge to the positive common value. Moreover, as shown in Fig. 6e and f, the total power mismatch is compensated by the BESS output power that is calculated by (\ref{eq:8.5}), and the steady-state frequency of the system restore to the nominal value accordingly; In the third stage, the total load reduces to $40$kW, the total power mismatch then changes from $-20$kW to $10$kW. As a result presented in Fig. 6b and c, the consensus value $\lambda^*$ decrease to $-0.14$, the BESSs switch from discharging mode to charging mode with SoC increasing, and the steady-state frequency maintains $50$Hz; The all the nominal active power $P_{\mathcal{R},i}^{\mathrm{nom}}$ increase $120\%$ and decrease $50\%$ in forth and fifth stage respectively. The corresponding tendency of SoC and the charging/discharging power of BESSs are represented in Fig. 6c and d. Fig. 6e shows that the controllers can always restore the frequency of the system to the nominal value copying with power of DGs and load changing.

The event intervals are depicted in Fig. 7, which implies that the communication burden is reduced in the context of ensuring the optimization purpose. Moreover, Fig. 7 shows that all the event intervals under event-triggered mechanism are positive, which demonstrates the Zeno behavior is efficiently avoided.

%%%%%%%%%%%%%%%%%%%%%%%%%%%%%%%%%%%%%%%%%%%%%%%%%%%%%%  Table I

\begin{figure}[!t]
\centering
\includegraphics[width=0.5\textwidth,height=2.8cm]{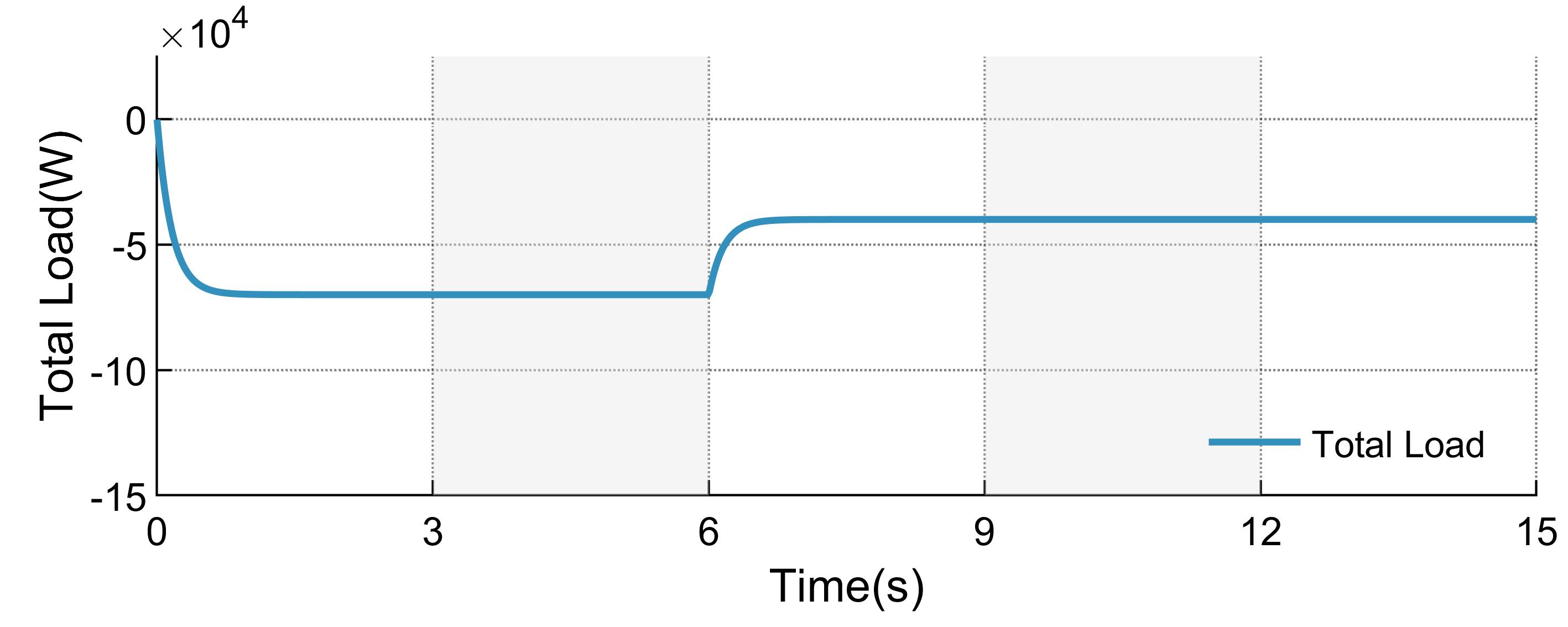}\\ \footnotesize {~~~~~~(a)}\\
\includegraphics[width=0.5\textwidth,height=2.8cm]{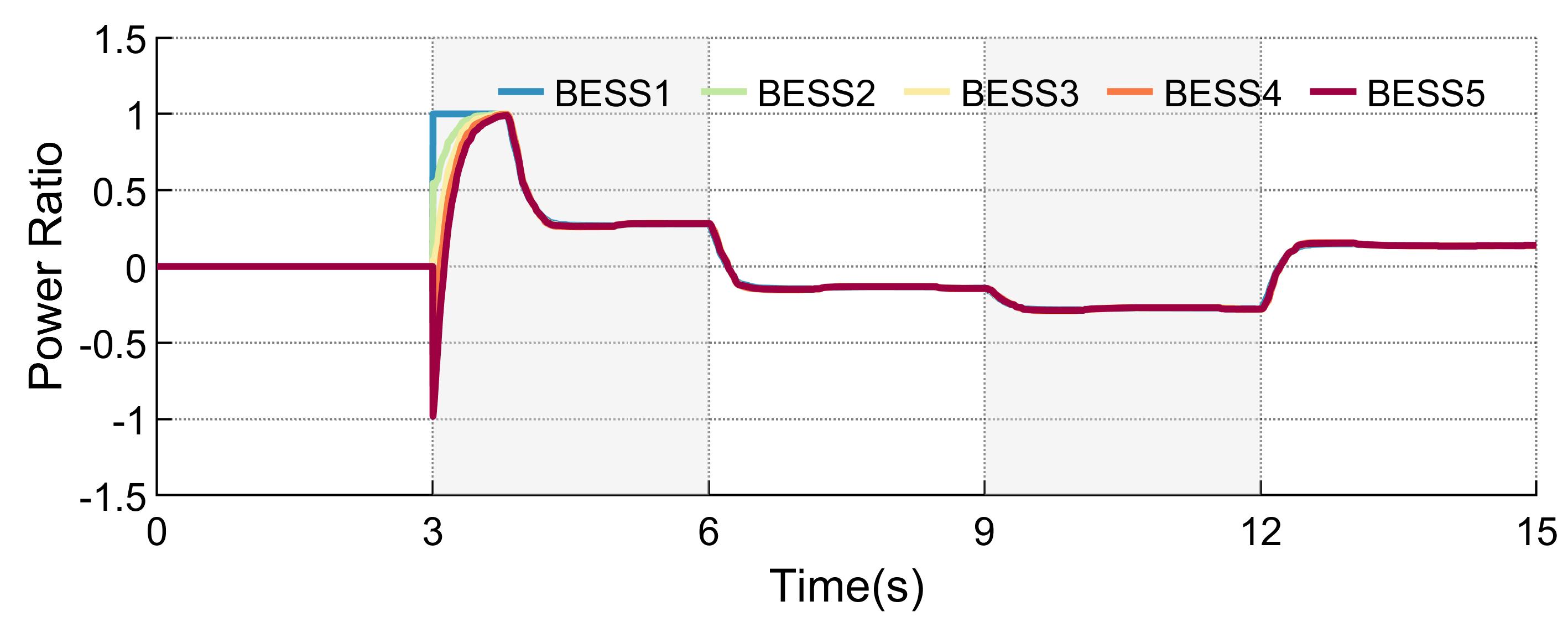}\\ \footnotesize {~~~~~~(b)}\\
\includegraphics[width=0.5\textwidth,height=2.8cm]{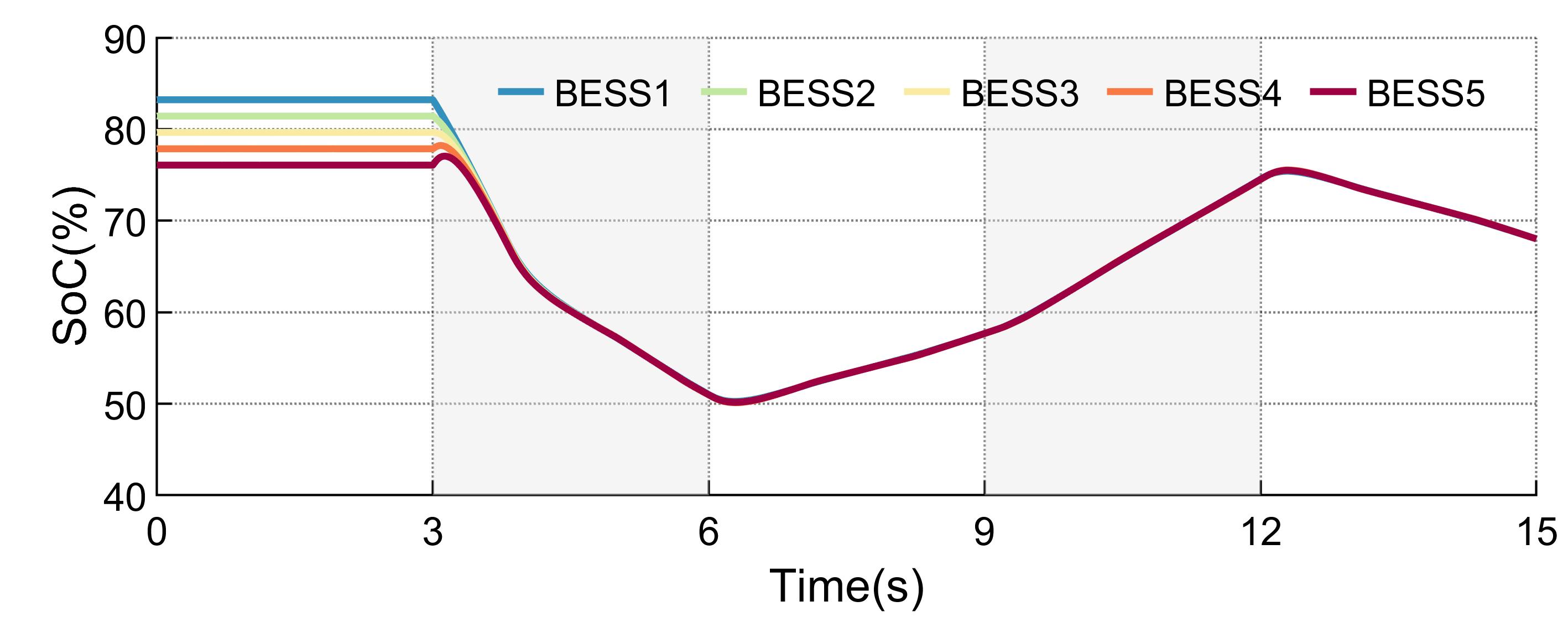}\\ \footnotesize {~~~~~~(c)}\\
\includegraphics[width=0.5\textwidth,height=2.8cm]{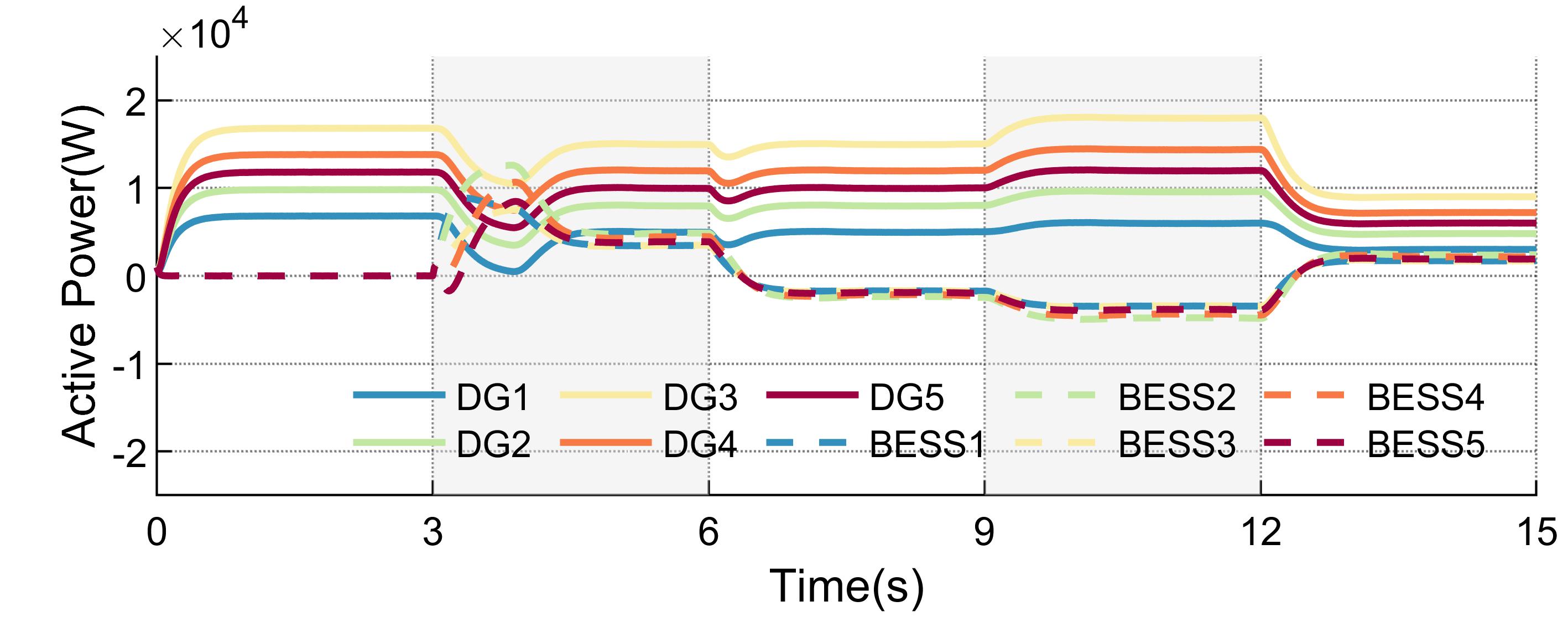}\\ \footnotesize {~~~~~~(d)}\\
\includegraphics[width=0.5\textwidth,height=2.8cm]{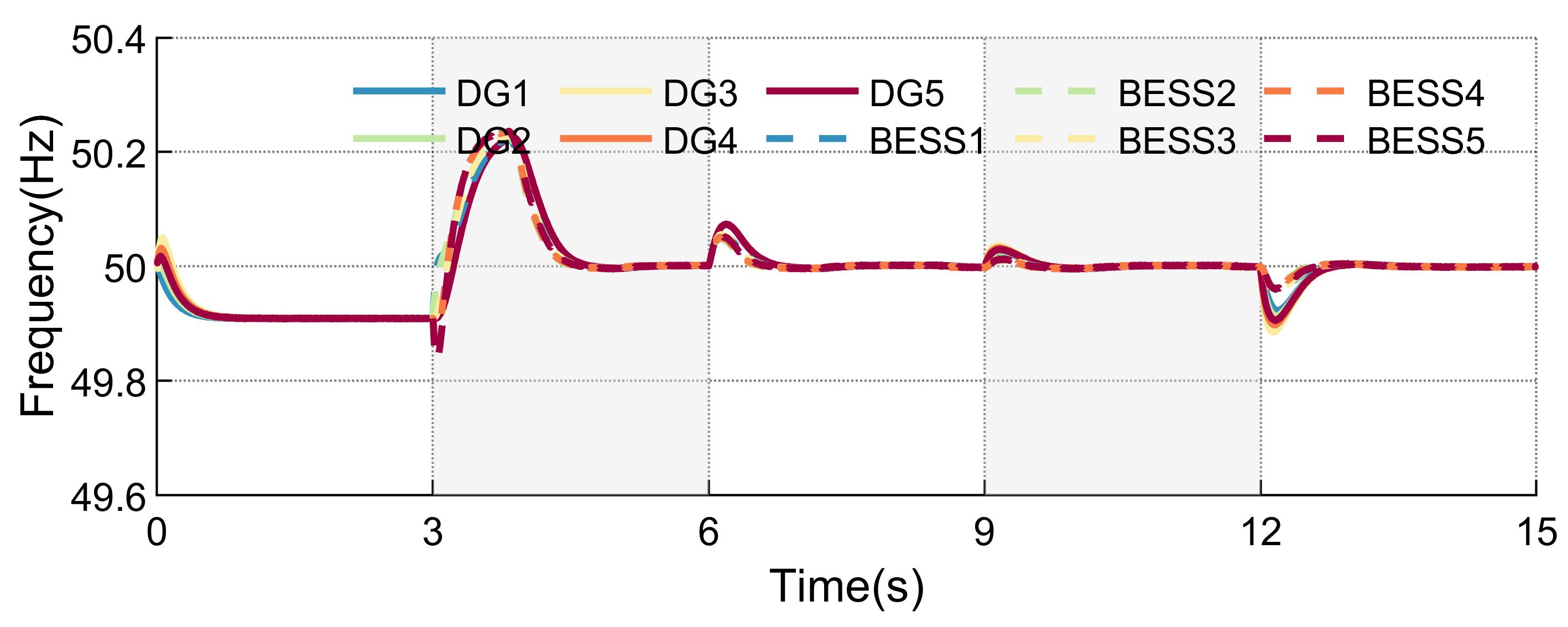}\\ \footnotesize {~~~~~~(e)}\\
\caption{Control performance without local constraints. (a) total load, (b) power ratios of BESSs, (c) SoC of BESSs, (d) output active power of DGs and BESSs, and (e) frequencies of DGs and BESSs.}\label{fig:6}
\end{figure}

\begin{figure}[!t]
\centering
\includegraphics[width=0.5\textwidth,height=2.8cm]{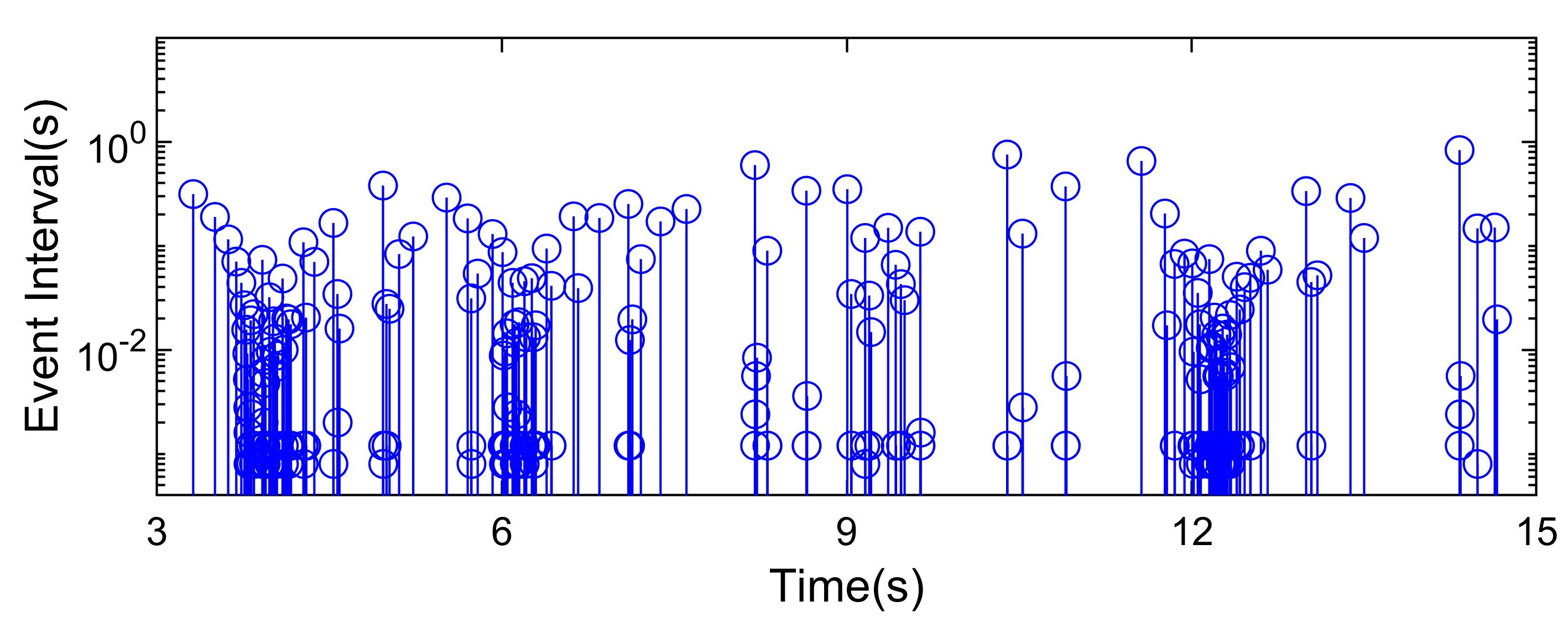}\\ \footnotesize {~~~~~~(a)}\\
\includegraphics[width=0.5\textwidth,height=2.8cm]{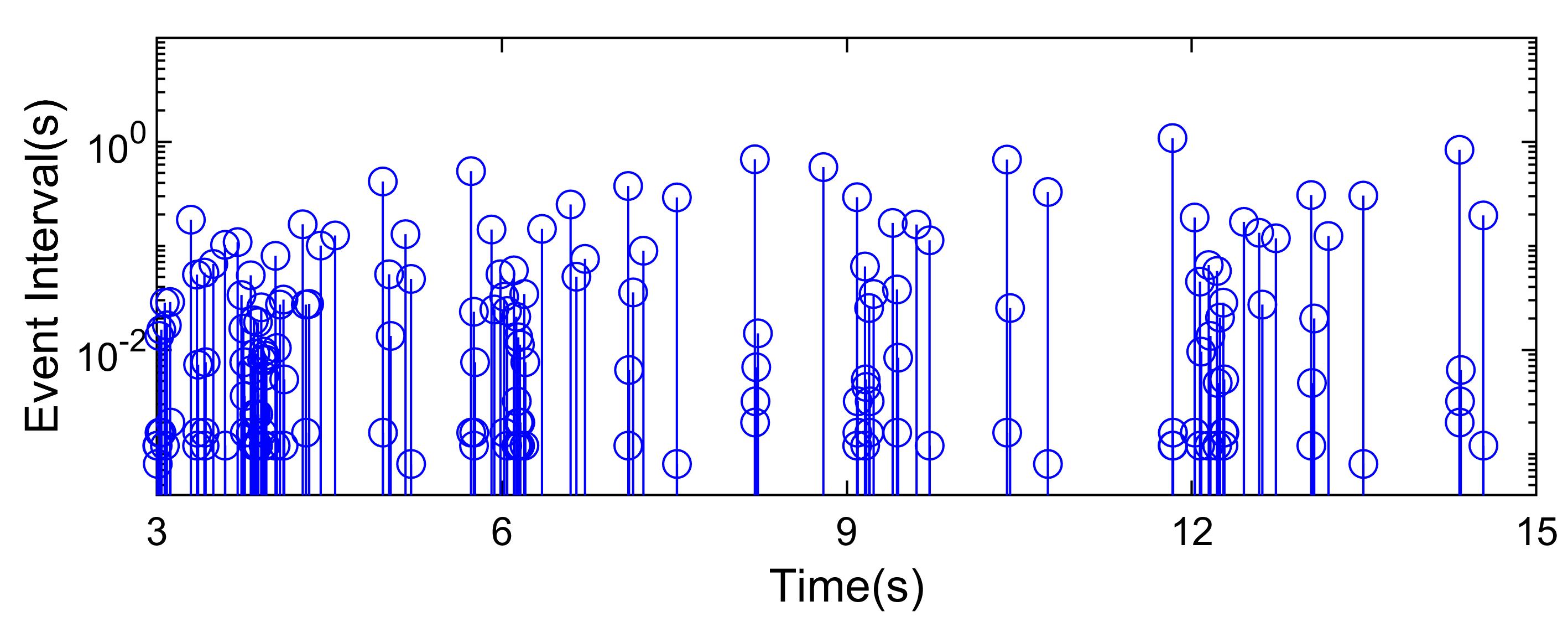}\\ \footnotesize {~~~~~~(b)}\\
\includegraphics[width=0.5\textwidth,height=2.8cm]{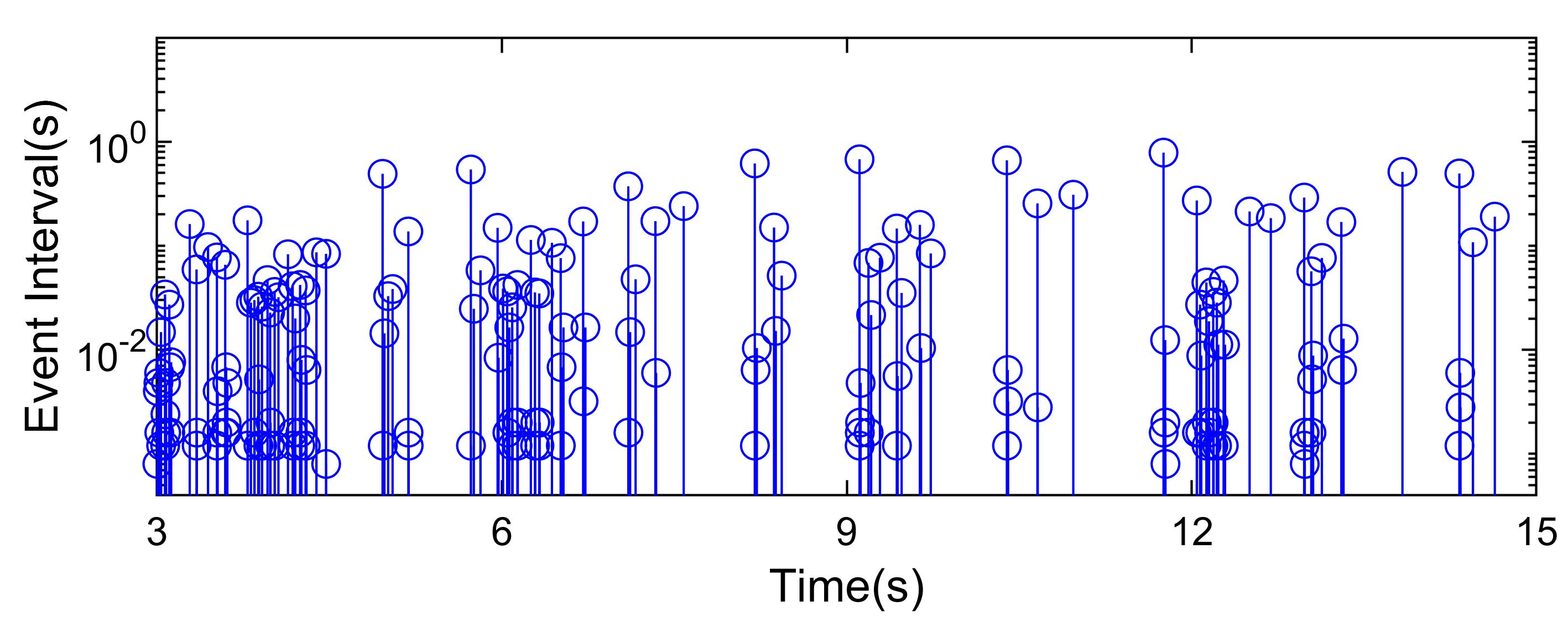}\\ \footnotesize {~~~~~~(c)}\\
\includegraphics[width=0.5\textwidth,height=2.8cm]{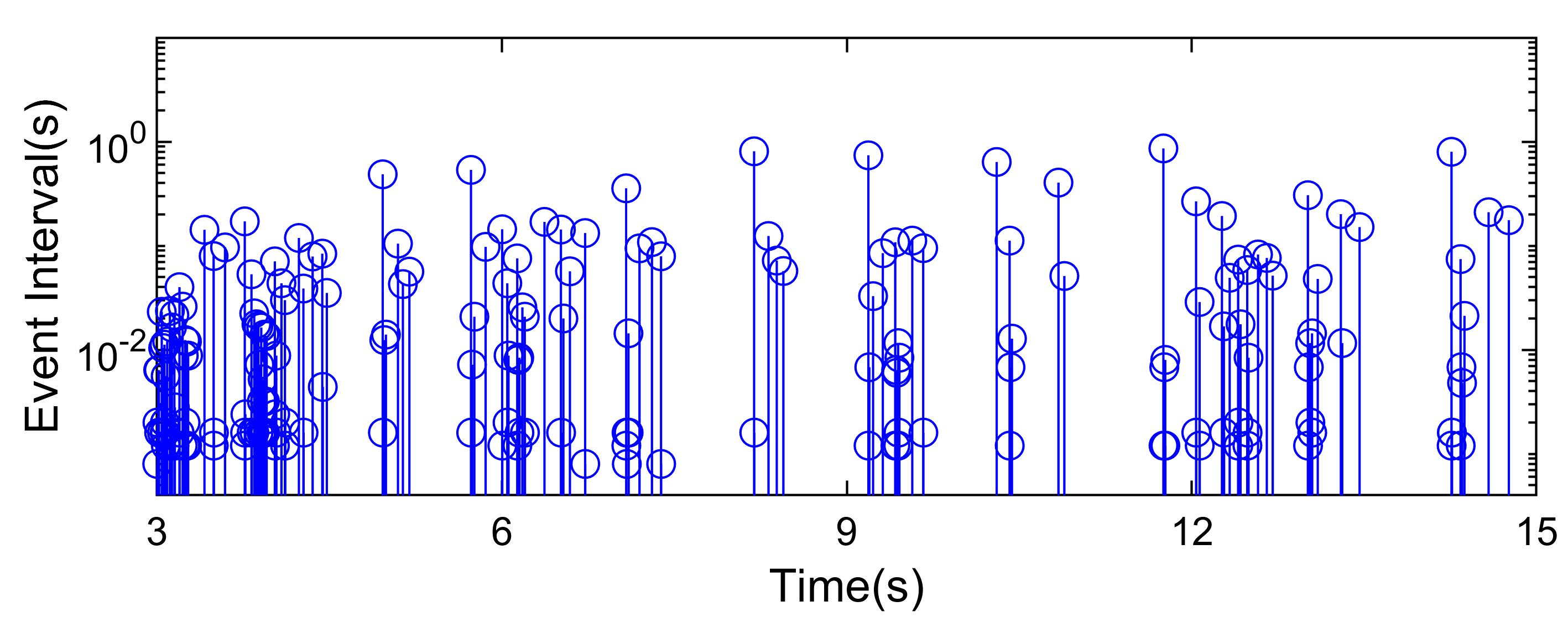}\\ \footnotesize {~~~~~~(d)}\\
\includegraphics[width=0.5\textwidth,height=2.8cm]{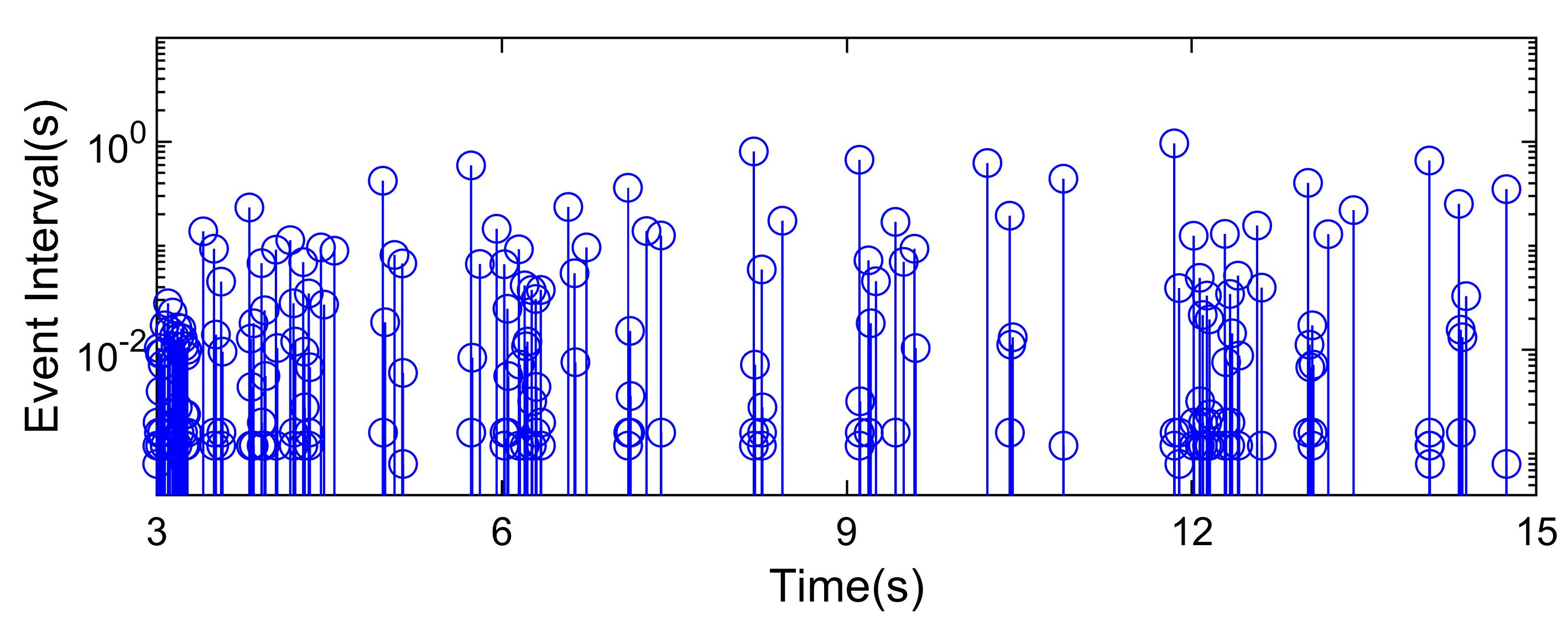}\\ \footnotesize {~~~~~~(e)}\\
\caption{Event instants and intervals of BESSs. (a)-(e) Event instants and intervals of BESSs 1-5.}\label{fig:7}
\end{figure}

\emph{Case 2: With local constraints}

In this case, the Soc and frequency deviation constraint considered. If the steady-state frequency of the system is in a acceptable range ($\pm$0.1Hz), the controller will not be activated until the steady-state frequency exceed its limitation. As shown in Fig. 8a, the total changes from $50$kW to $58.5$kW in the first $30$s, and then reduce to $39$kW from $t=3$s to $t=6$s. The load change causes the total power mismatch varying from $-8.5$kW to $11$kW, and the steady-state frequency ranging from $49.96$ to $50.05$Hz, as shown in Fig. 8e. After $t=6$s, the total load keeps rising, and the steady-state frequency reaches its lower bound, $49.9$Hz, at $t=10$s. At the same time, the controller is activated and the BESSs output active power to compensate the total power mismatch. As a result depicted in Fig. 8b-d, $\lambda_i(t)$ and the active power of BESSs keeps increasing, and their SoC decreases accordingly. Then, the steady-state frequency is restored to and maintained at $50$Hz, as shown in Fig. 8e. Fig. 8f presents the event intervals of BESS 3 after $t=10$. It is easily seen that this constraint effectively reduces the charging/discharging switching frequency caused by load/generation jitter. Note that all the SoC of BESSs at $t=15$s approaches to $30$\%, $\lambda_i(t)$ and the charging/discharging power of BESSs will be set to zero if the SoC falls to $30$\%.
\begin{figure}[!t]
\centering
\includegraphics[width=0.5\textwidth,height=2.8cm]{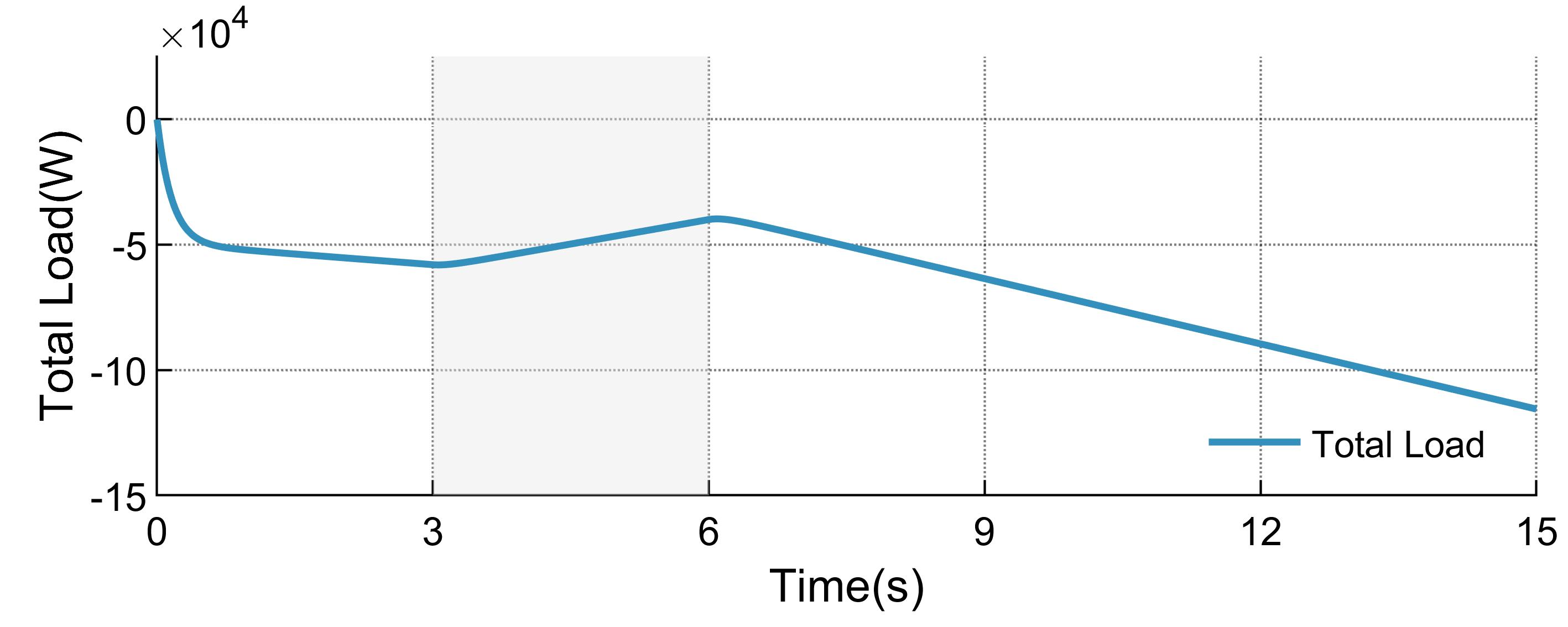}\\ \footnotesize {~~~~~~(a)}\\
\includegraphics[width=0.5\textwidth,height=2.8cm]{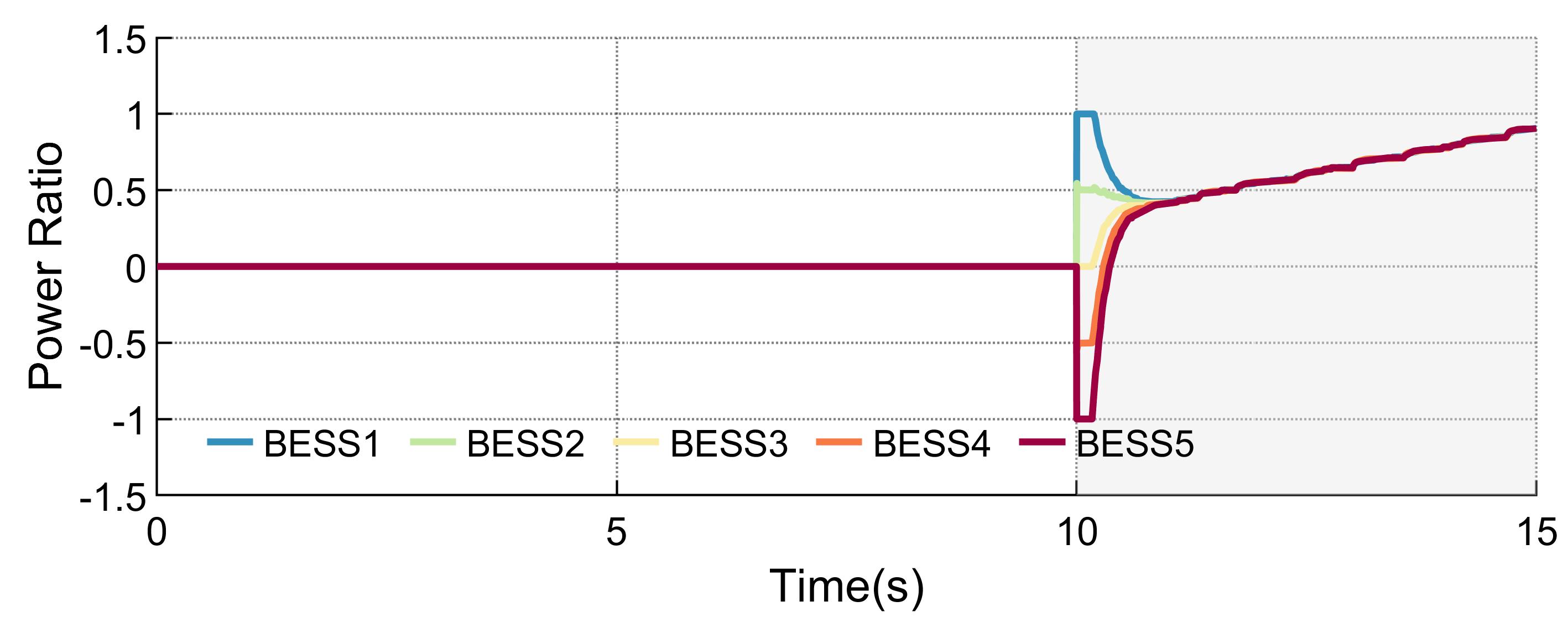}\\ \footnotesize {~~~~~~(b)}\\
\includegraphics[width=0.5\textwidth,height=2.8cm]{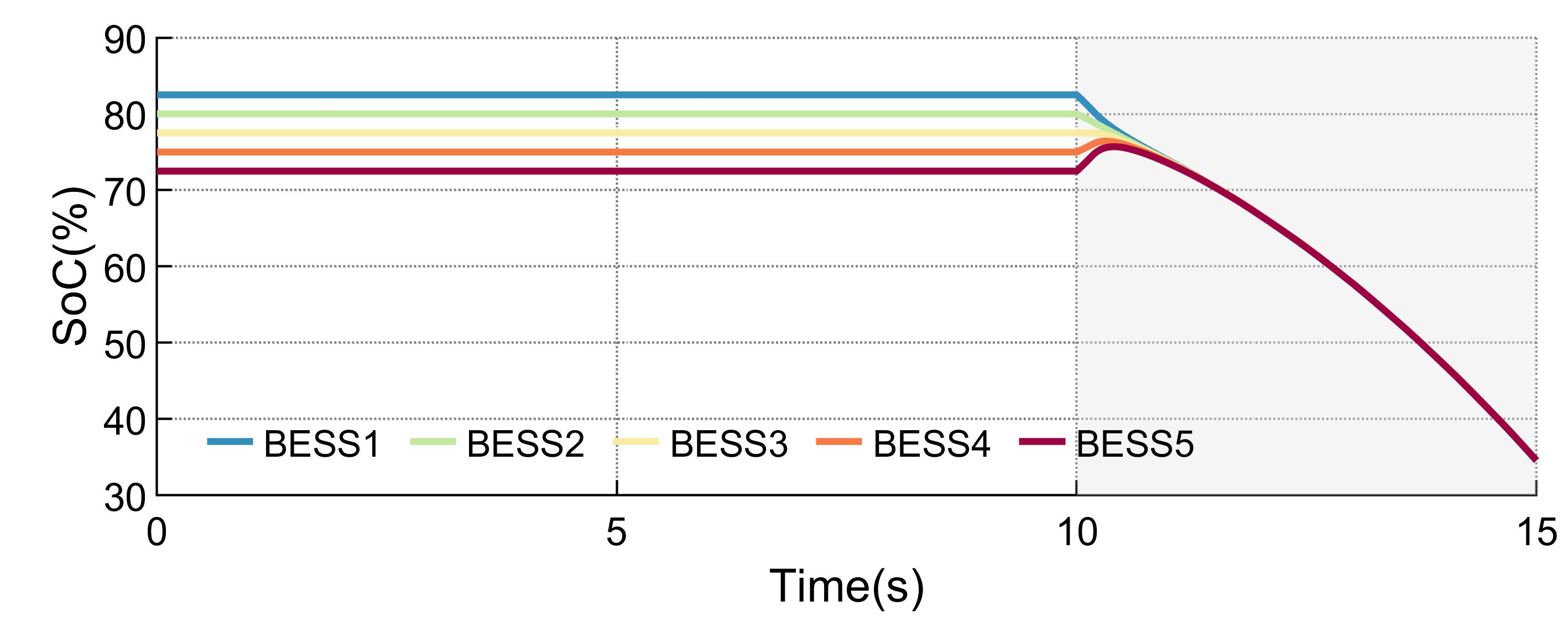}\\ \footnotesize {~~~~~~(c)}\\
\includegraphics[width=0.5\textwidth,height=2.8cm]{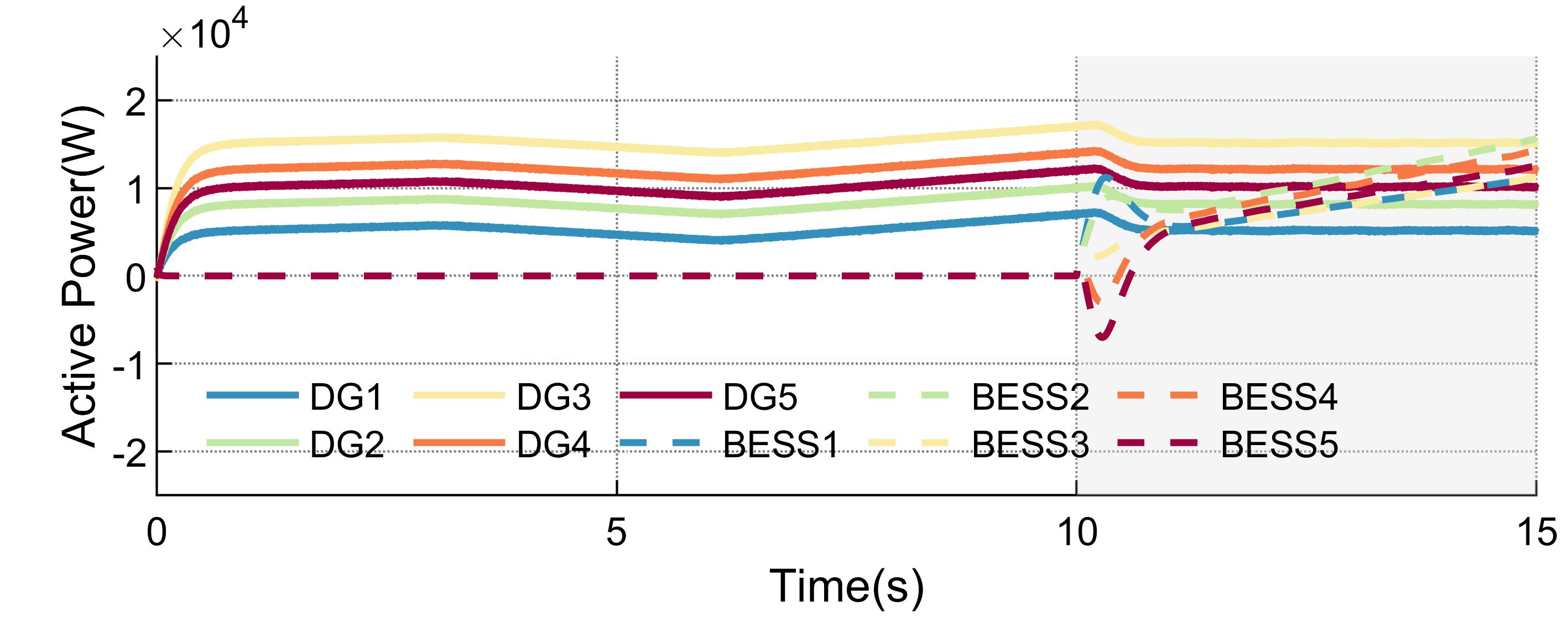}\\ \footnotesize {~~~~~~(d)}\\
\includegraphics[width=0.5\textwidth,height=2.8cm]{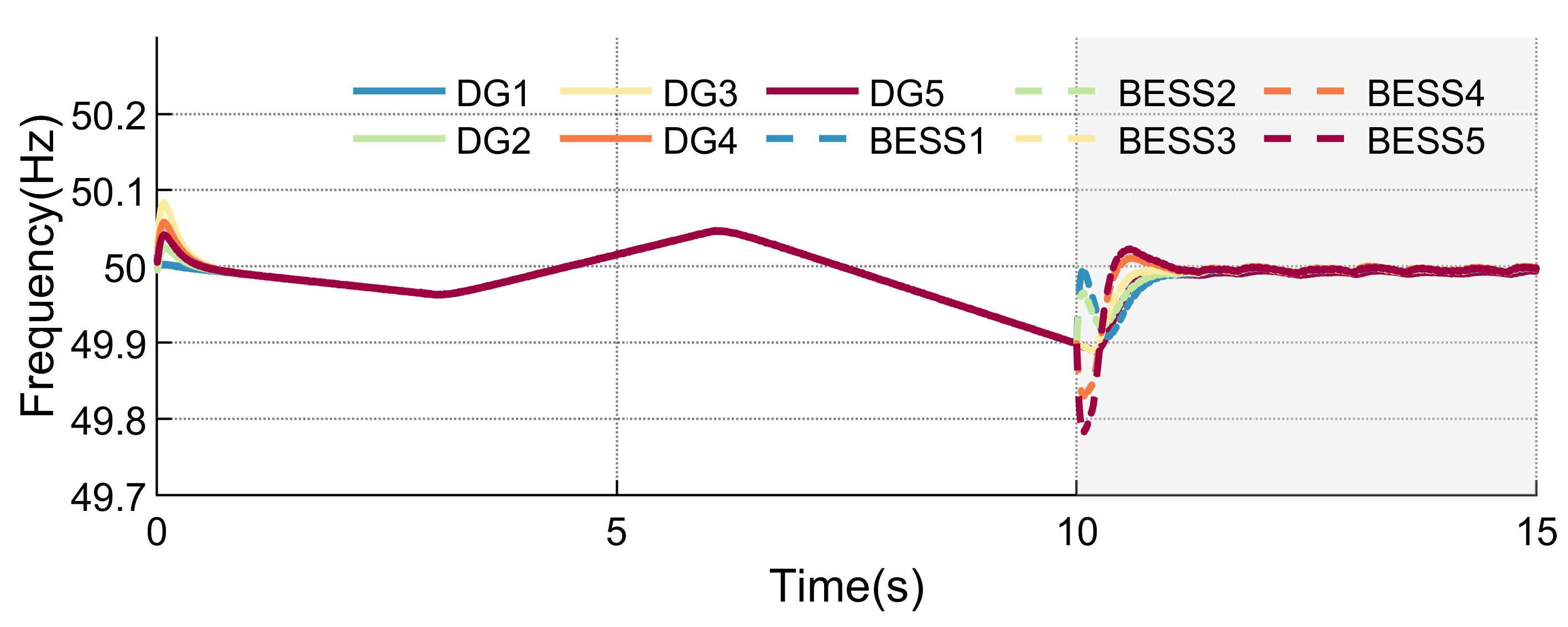}\\ \footnotesize {~~~~~~(e)}\\
\includegraphics[width=0.5\textwidth,height=2.8cm]{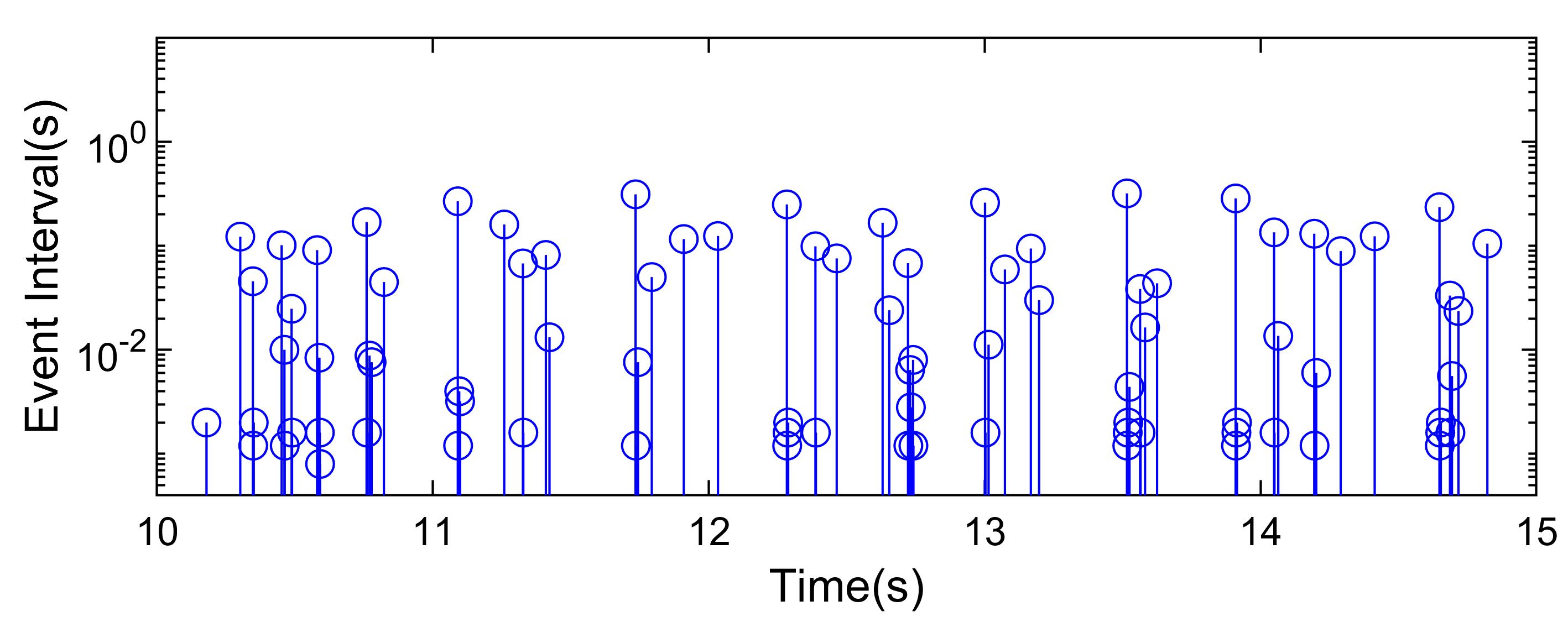}\\ \footnotesize {~~~~~~(f)}\\
\caption{Control performance with local constraints. (a) total load, (b) power ratios of BESSs, (c) SoC of BESSs, (d) output active power of DGs and BESSs, (e) frequencies of DGs and BESSs, and (f) event instants and intervals of BESS 3.}\label{fig:8}
\end{figure}

\begin{table}[t]
\renewcommand\arraystretch{1.5}
  \centering
  \caption{Convergence Times Of Several Methods.}
  \setlength{\tabcolsep}{3pt}
    \begin{tabular}{|p{50pt}<{\centering}|p{100pt}<{\centering}|p{100pt}<{\centering}|p{100pt}<{\centering}|}
    \hline
    &Asymptotic controller&Finite controller &Improved finite-time controller\\
    \hline
    \hline
    $\tau_1$&2.72s&2.84s&6.63s \\ \hline
    $\tau_2$&2.74s&11.67s&14.21s \\ \hline
    \end{tabular}
\end{table}

\emph{Case 3: Comparison with other methods}

\begin{figure}[!t]
\centering
\includegraphics[width=0.5\textwidth,height=2.8cm]{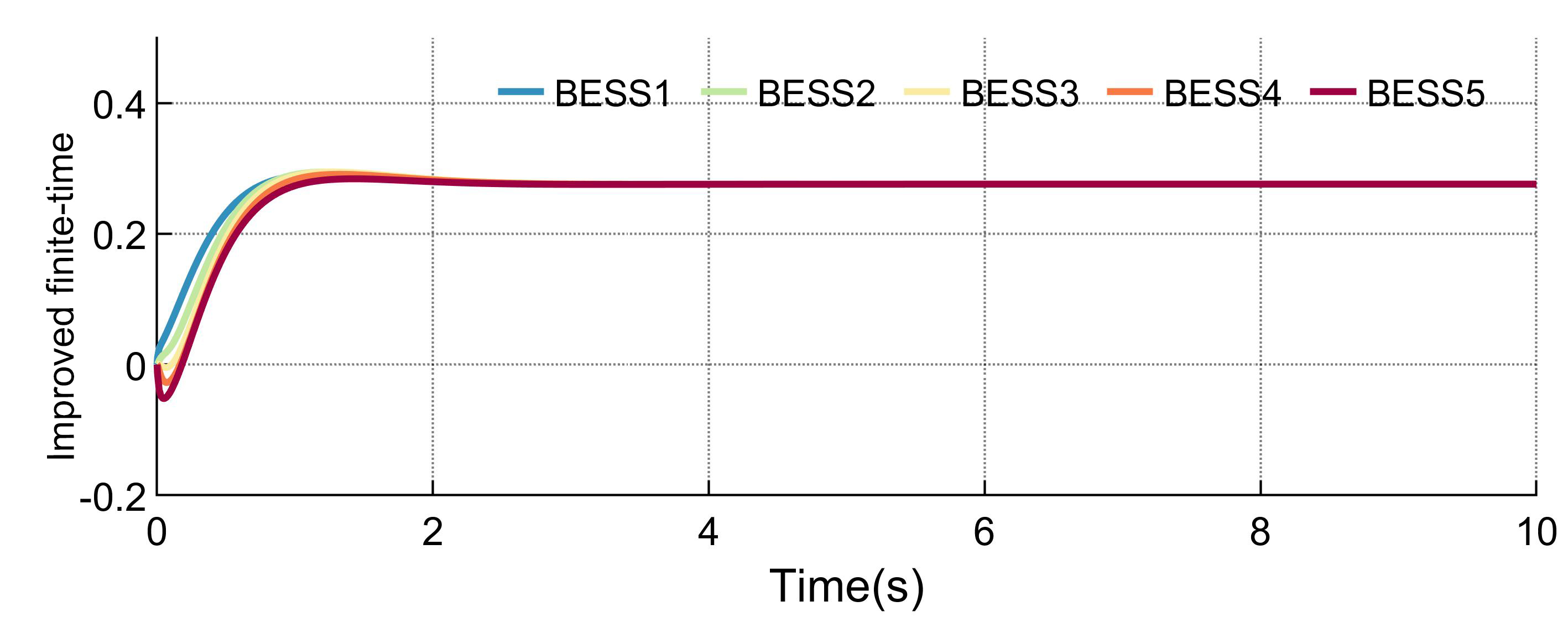}\\
\includegraphics[width=0.5\textwidth,height=2.8cm]{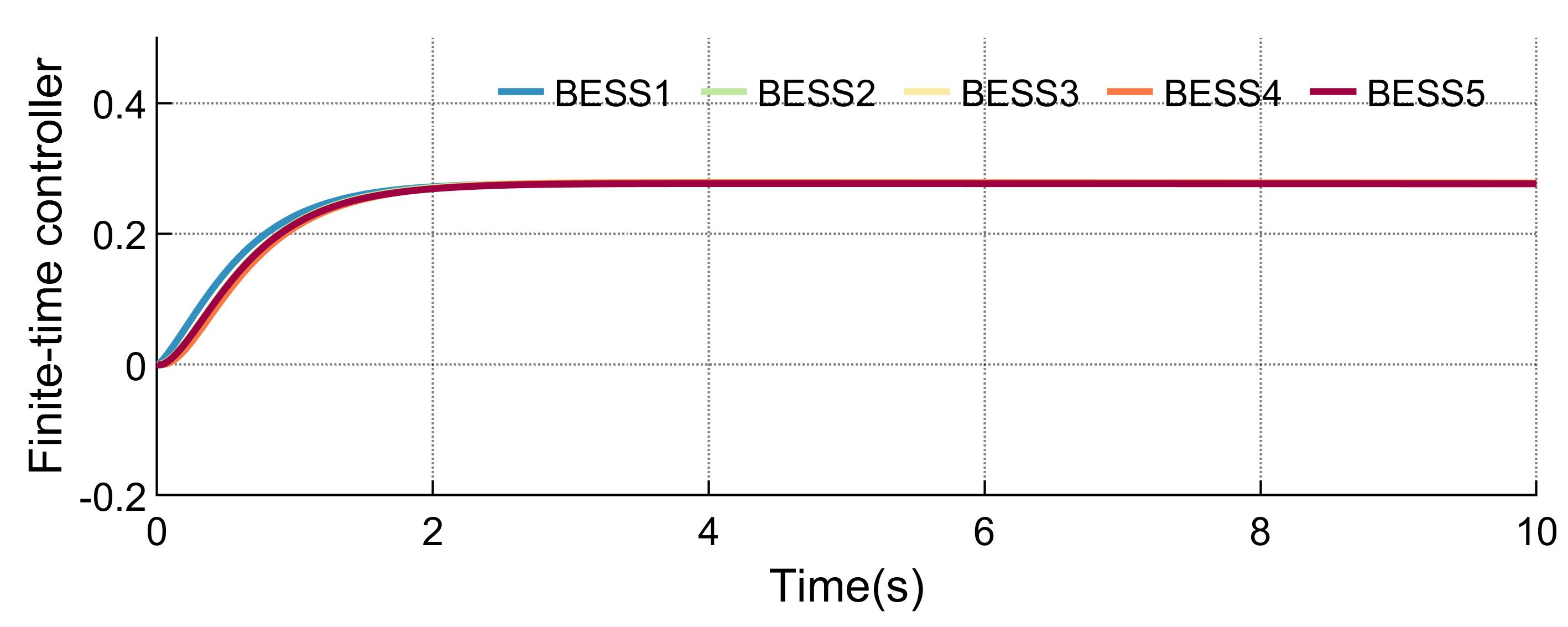}\\
\includegraphics[width=0.5\textwidth,height=2.8cm]{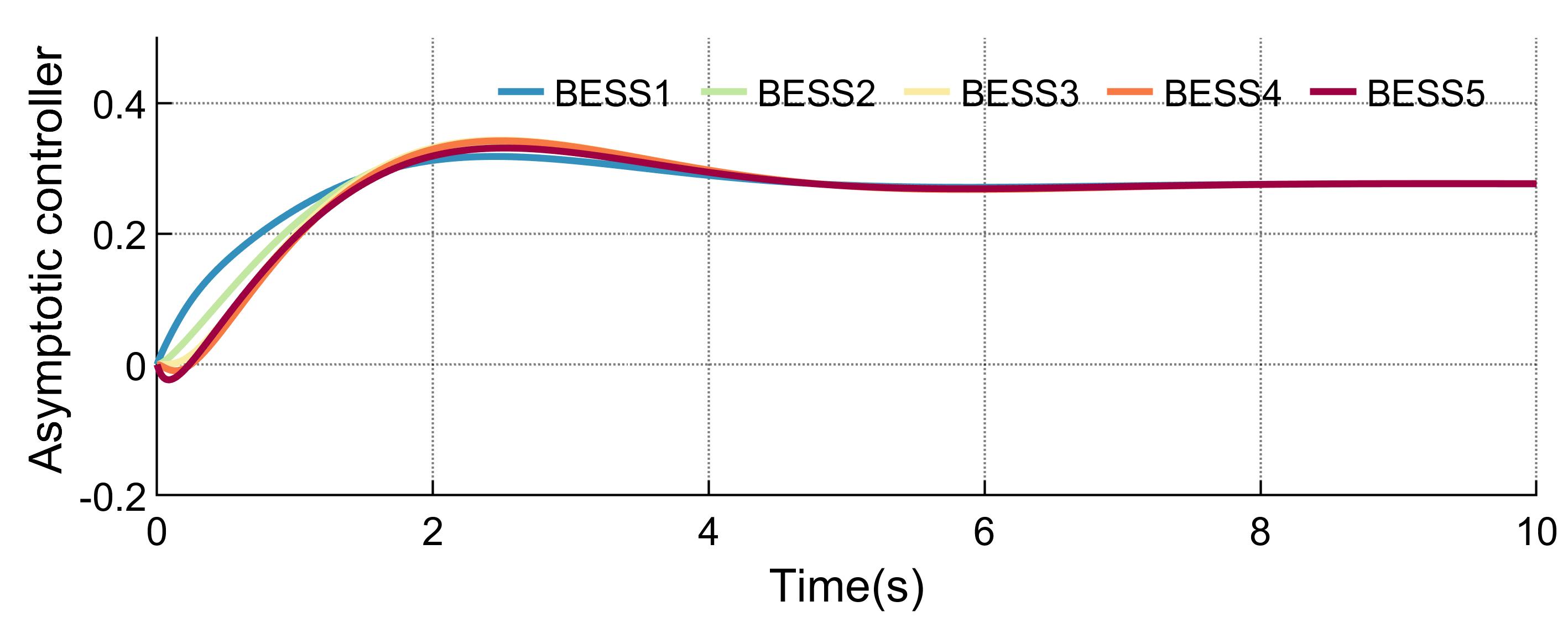}\\ \footnotesize {~~~~~~(a)}\\
\includegraphics[width=0.5\textwidth,height=2.8cm]{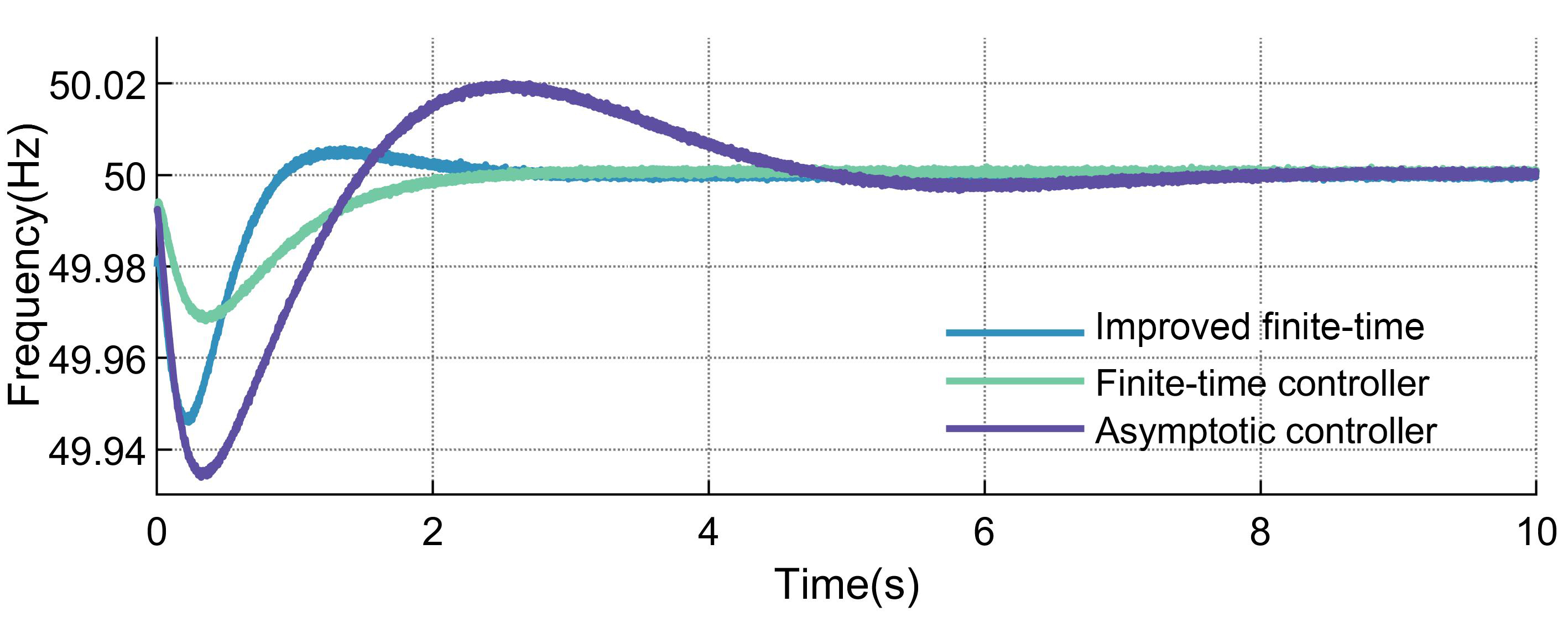}\\ \footnotesize {~~~~~~(b)}\\
\caption{Control performance without charging rate constraints of BESSs. (a) evolution of incremental costs, (b) charging rates, (c) frequency response of inverters associated with all BESSs, and (d) total power balance.}\label{fig:9}
\end{figure}
\begin{figure}[!t]
\centering
\includegraphics[width=0.5\textwidth,height=2.8cm]{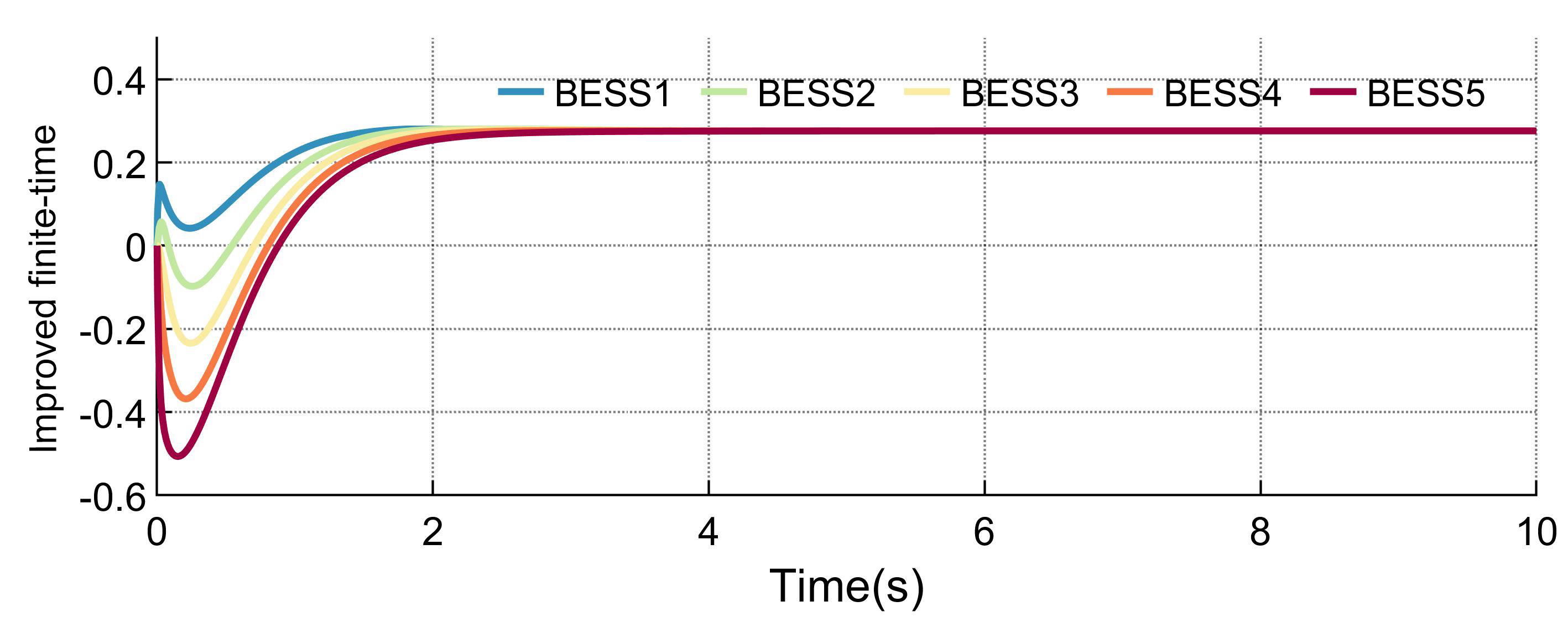}\\
\includegraphics[width=0.5\textwidth,height=2.8cm]{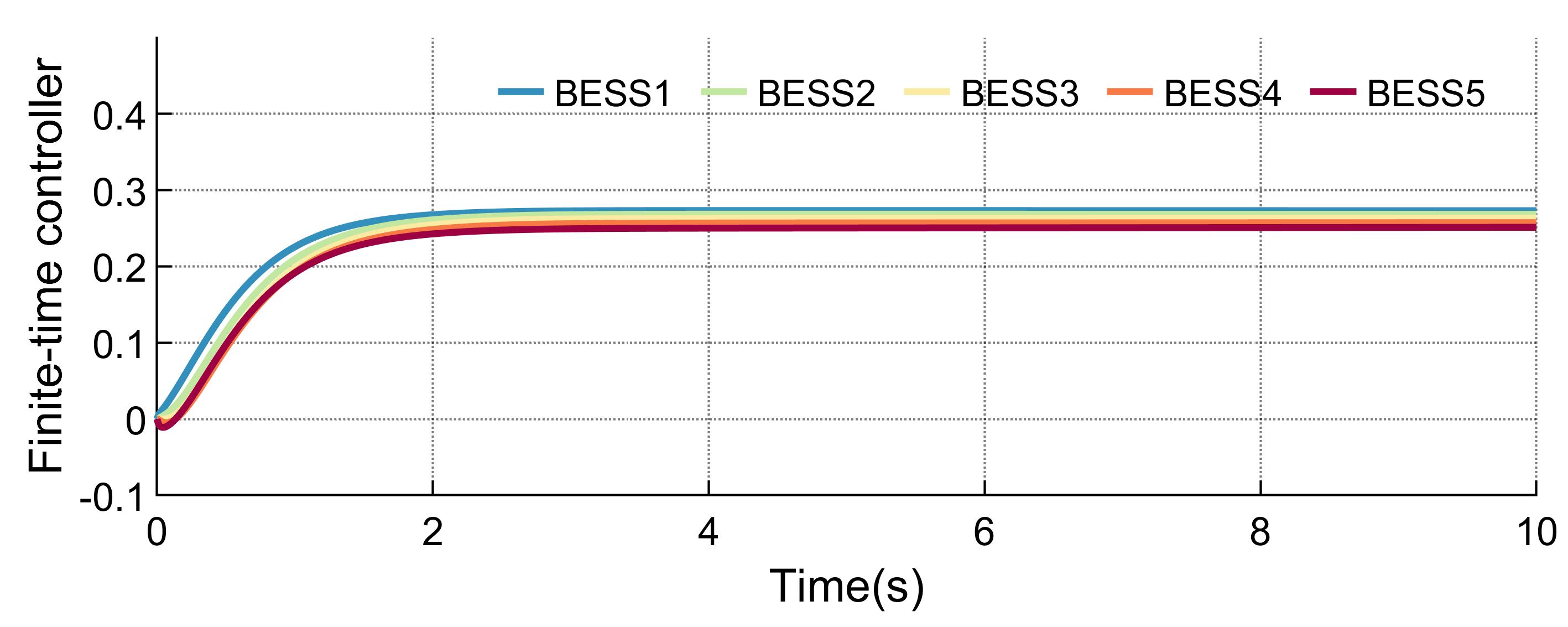}\\
\includegraphics[width=0.5\textwidth,height=2.8cm]{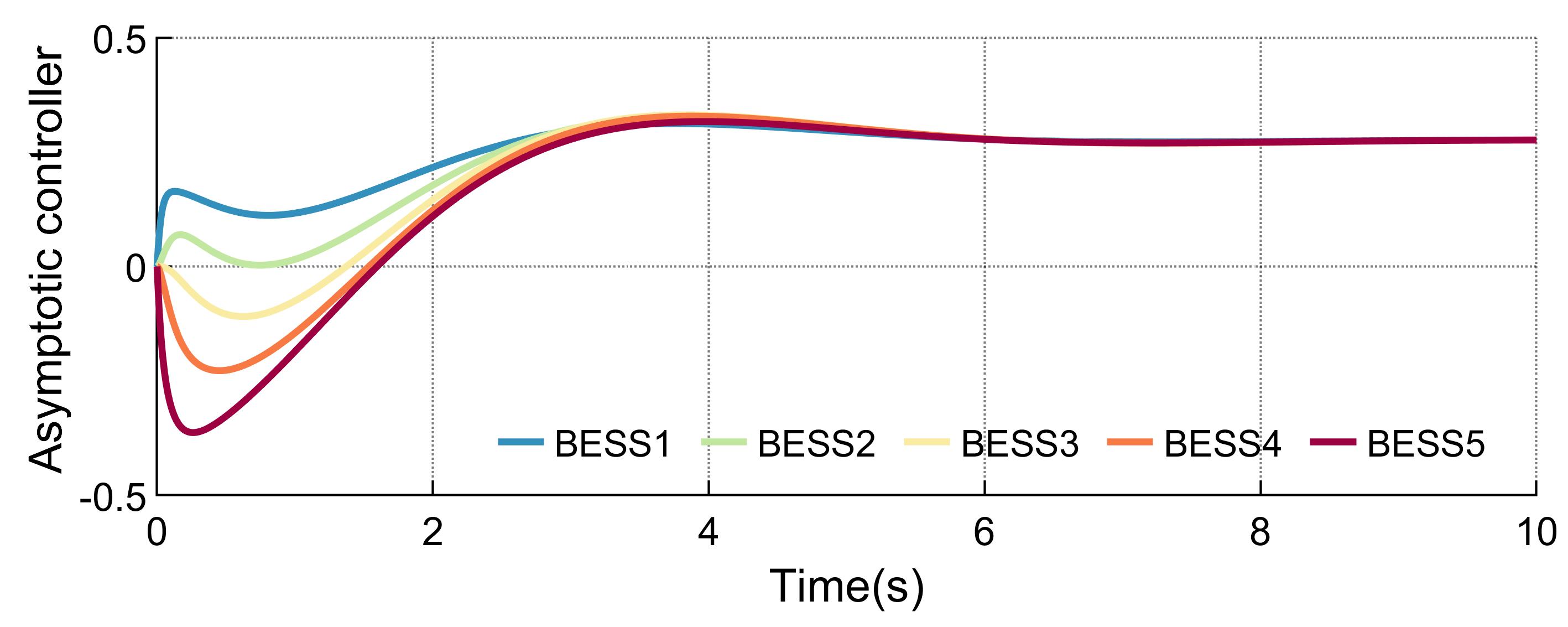}\\ \footnotesize {~~~~~~(a)}\\
\includegraphics[width=0.5\textwidth,height=2.8cm]{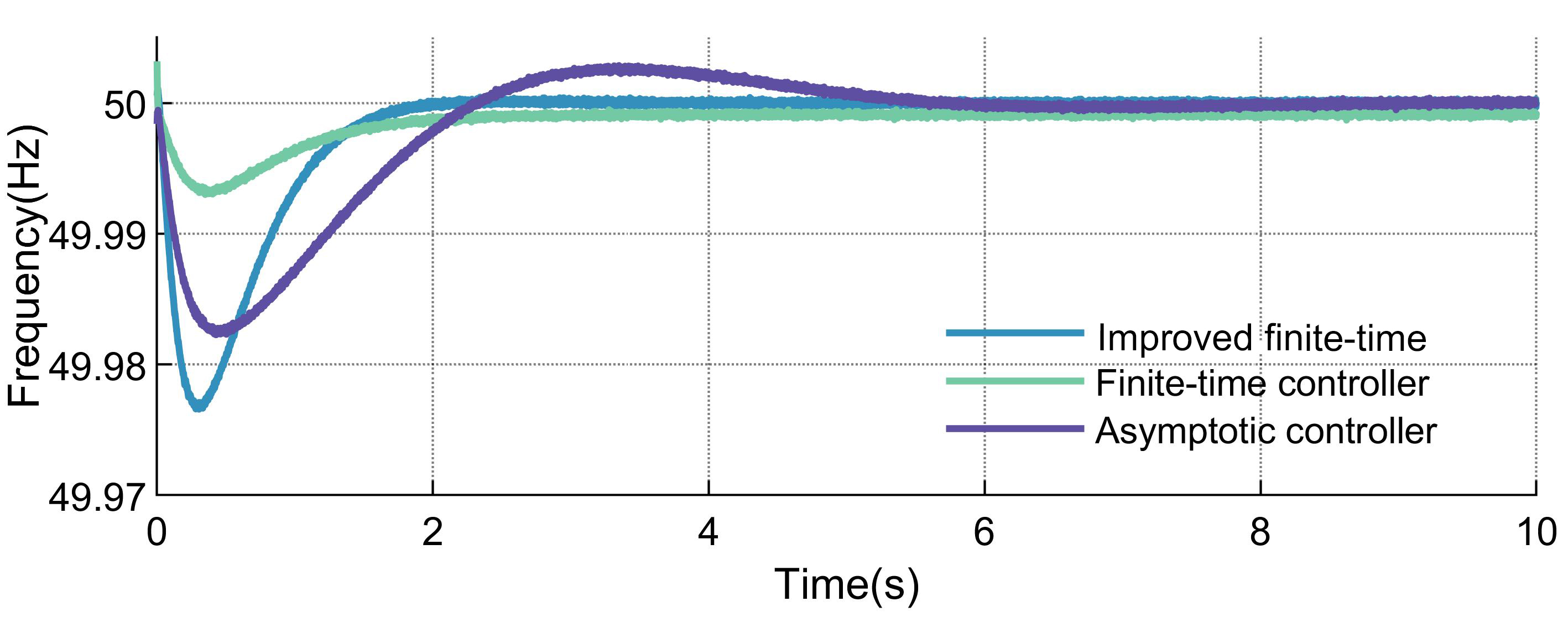}\\ \footnotesize {~~~~~~(b)}\\
\caption{Control performance without charging rate constraints of BESSs. (a) evolution of incremental costs, (b) charging rates, (c) frequency response of inverters associated with all BESSs, and (d) total power balance.}\label{fig:10}
\end{figure}

{\color{blue}The proposed method is compared with the finite-time controller in \cite{12,13} and the conventional asymptotic controller in \cite{191} with the same parameters, in this case. For a given initial condition, the dynamic responses copying with a load change from $0$ to $70$kW of $\lambda_i$ and the frequency curves, under the above three controllers, are obtained. Then, the initial condition is expanded and the obtained results were compared to the previous ones. As Fig. 10a shown, all $\lambda_i(t)$ reach consensus within a settling time $2.26$s using the proposed improved finite-time controller, while they reach consensus within a settling time $5.72$s using the finite-time controller. Although $\lambda_i(t)$ can not be equal to the consensus value in a finite time with the asymptotic controller, they reach the steady state with an acceptable error within $8.25$s. It can be obtained in Fig. 10b that the proposed finite-time controller enforces the frequency restore to $50$Hz within the shortest time. After expanding the initial condition, Fig.11a shows that the proposed controller has a nearly constant settling time, while the settling time of the finite-time controller increases greatly. Moreover, the convergence speed of the asymptotic controller decreases. As a result in Fig.11b, the frequency restoration speed of the proposed controller remains at high level regardless of the variation of initial condition. The comparison results are shown in Table III, in which, $\tau_1$ and $\tau_2$ represent the convergence time for two initial conditions (for asymptotic controller, the time calculated by the approximate stable state).}

\section{Conclusion}
{\color{blue}Distributed finite-time frequency restoration and SoC balancing problems in islanded ac microgrids with an undirected connected communication network, are addressed. The control protocol, under a event-triggered mechanism, is designed to share the active power among BESSs based on their power ratings, and the total power mismatch of the system is compensated such that restore the steady-state frequency is enabled to restore to the nominal value. Lyapunov proofs and  homogeneous approximation theory establish the stability and the upper bounds for convergence time of the proposed secondary controller. Numerical simulations show that, compared to the conventional finite-time and asymptotic controller, the proposed event-triggered finite-time scheme improves the synchronization speed, with a guaranteed bound on the convergence time that does not change with initial conditions, and achieves significantly reduces communication burden. Future works may focus on the situations considering the communication delay and cyber security problems.}

\ifCLASSOPTIONcaptionsoff
  \newpage
\fi

\end{document}